\documentclass[reprint,aps,showpacs,prx,superscriptaddress,floatfix,10pt,nofootinbib]{revtex4-2}
\usepackage[T1]{fontenc}
\usepackage[utf8]{inputenc}
\usepackage{amsmath}
\usepackage{amssymb}
\usepackage{esint}
\usepackage{graphicx}
\usepackage{bbold}
\usepackage{appendix}
\usepackage{bibunits}
\usepackage{verbatim}
\usepackage{bm}
\usepackage{xr-hyper}
\usepackage{lipsum} 
\defaultbibliographystyle{myapsrev4-2}
\makeatletter
\usepackage{color,xcolor}
\definecolor{deepgreen}{rgb}{0.0, 0.5, 0.0}

\usepackage{physics} 
\usepackage{xcolor} 

\definecolor{mygreineren}{RGB}{0, 126, 0}
\definecolor{myorange}{RGB}{255, 136, 19}
\definecolor{mymagenta}{RGB}{200, 0, 100}

\def\@fnsymbol#1{\ensuremath{\ifcase#1\or \dagger\or \ddagger\or
   \mathsection\or \mathparagraph\or \|\or **\or \dagger\dagger
   \or \ddagger\ddagger \else\@ctrerr\fi}}
    \makeatother

\newcommand{\jqi}{
Joint Quantum Institute and Department of Physics, University of Maryland, College Park, MD 20742, USA
}
\newcommand{\dqc}{
Duke Quantum Center and Department of Physics, Duke University, Durham, NC 27701, USA
}

\makeatother

\begin{document}
\title{The phase diagram of quantum chromodynamics \\
in one dimension on a quantum computer}
% \author{}
% \affiliation{}

\author{Anton T. Than*}
\affiliation{\jqi}

\author{Yasar Y. Atas*}
\affiliation{Institute for Quantum Computing, University of Waterloo, Waterloo, ON, Canada, N2L 3G1}
\affiliation{Department of Physics \& Astronomy, University of Waterloo, Waterloo, ON, Canada, N2L 3G1}
\author{Abhijit Chakraborty*}
\affiliation{Institute for Quantum Computing, University of Waterloo, Waterloo, ON, Canada, N2L 3G1}
\affiliation{Department of Physics \& Astronomy, University of Waterloo, Waterloo, ON, Canada, N2L 3G1}

\author{Jinglei Zhang}
\affiliation{Institute for Quantum Computing, University of Waterloo, Waterloo, ON, Canada, N2L 3G1}
\affiliation{Department of Physics \& Astronomy, University of Waterloo, Waterloo, ON, Canada, N2L 3G1}

\author{\mbox{Matthew T. Diaz}}
\affiliation{\jqi}

\author{Kalea Wen}
\affiliation{College of William \& Mary,
Williamsburg, VA 23187, USA}
\affiliation{\jqi}

\author{Xingxin Liu}
\affiliation{\jqi}

\author{Randy Lewis}
 \affiliation{Department of Physics and Astronomy, York University, Toronto, ON, Canada, M3J 1P3}

\author{Alaina M. Green$^{\dagger}$}
\affiliation{\jqi}

\author{Christine A. Muschik$^{\dagger}$}
\affiliation{Institute for Quantum Computing, University of Waterloo, Waterloo, ON, Canada, N2L 3G1}
\affiliation{Department of Physics \& Astronomy, University of Waterloo, Waterloo, ON, Canada, N2L 3G1}
\affiliation{Perimeter Institute for Theoretical Physics, Waterloo, ON, Canada, N2L 2Y5}

\author{Norbert M. Linke$^{\dagger}$}
\affiliation{\jqi}
\affiliation{\dqc}

\date{\today}

%\maketitle
\def\thefootnote{*}\footnotetext{These authors contributed equally to this work.}

\def\thefootnote{$\dagger$}\footnotetext{These authors jointly supervised this work.}

\begin{bibunit}

% \begin{abstract}
% The quantum chromodynamics (QCD) phase diagram is key to answering open questions in physics, from states of matter in neutron stars to the early universe \cite{guenther2021overview}. However, classical simulations of QCD face significant computational barriers, such as the sign problem at finite matter densities \cite{nagata2022finite}. Quantum computing offers a promising solution to overcome these challenges \cite{bauer2023quantum,banuls2020review,di2024quantum,humble2022snowmass}. Here, we take an important step toward exploring the QCD phase diagram with quantum devices by preparing thermal states in one-dimensional non-Abelian gauge theories. We experimentally simulate the thermal states of SU(2) and SU(3) gauge theories at finite densities on a trapped-ion quantum computer using a variational method. This is achieved by introducing two  features: Firstly, we add motional ancillae to the existing qubit register to efficiently prepare thermal probability distributions. Secondly, we introduce gauge-invariant measurements to enforce local gauge symmetries. This work marks the first lattice gauge theory quantum simulation of QCD at finite density and temperature for two and three colors, laying the foundation to explore QCD phenomena on quantum platforms.
% \end{abstract}
% \maketitle

\begin{abstract}
\vspace{1ex}
\section*{Abstract} 
The quantum chromodynamics (QCD) phase diagram, which reveals the state of strongly interacting matter at different temperatures and densities, is key to answering open questions in physics, ranging from the behavior of particles in neutron stars to the conditions of the early universe. However, classical simulations of QCD face significant computational barriers, such as the sign problem at finite matter densities. Quantum computing offers a promising solution to overcome these challenges. Here, we take an important step toward exploring the QCD phase diagram with quantum devices by preparing thermal states in one-dimensional non-Abelian gauge theories. We experimentally simulate the thermal states of SU(2) and SU(3) gauge theories at finite densities on a trapped-ion quantum computer using a variational method. This is achieved by introducing two features: Firstly, we add motional ancillae to the existing qubit register to efficiently prepare thermal probability distributions. Secondly, we introduce charge-singlet measurements to enforce color-neutrality constraints. This work marks the first lattice gauge theory quantum simulation of QCD at finite density and temperature for two and three colors, laying the foundation to explore QCD phenomena on quantum platforms.
\end{abstract}
\maketitle

\section{Introduction} 
The phase diagram of quantum chromodynamics (QCD) underpins our understanding of possible phases of matter in nature and addresses foundational questions in nuclear physics, particle physics, and cosmology. The QCD phase diagram maps out quarks and gluons across various temperatures and densities (non-zero fermionic chemical potentials). Despite intense scientific interest in exploring the QCD phase diagram \cite{guenther2021overview,aarts2023phase,muroya2003lattice,karsch2003hadron,aoki2006order,aoki2006qcd,barbour1998results}, most current numerical approaches, which are based on Monte Carlo methods, are hindered by sign problems \cite{nagata2022finite}.
 
Several strategies have been explored for addressing the sign problem in order to probe different regions of the QCD phase diagram \cite{fodor2007density,fodor2002new,csikor2003qcd,allton2002qcd,alexandru2016monte,seiler2013gauge,ejiri2008canonical,guenther2021overview,alexandru2022complex,nagata2022finite,ratti2018lattice,achenbach2024present}. One promising approach to avoid sign problems at nonzero fermionic chemical potentials is to adopt the Hamiltonian formalism \cite{Wilson1974sk,gregory2000hamiltonian}. Initial studies using tensor-network states within this framework show encouraging results \cite{montangero2018introduction, silvi_finite-density_2017,magnifico2024tensor,banuls2023tensor,silvi2014lattice,rico2014tensor,banuls_efficient_2017,silvi_tensor_2019,magnifico2021lattice,cataldi2024simulating,pichler2016real,felser_two-dimensional_2020,banuls2020review,banuls2018tensor}.
Another path forward is the use of quantum computing, which offers a fundamentally sign-problem-free approach for future lattice gauge theory (LGT) calculations at nonzero densities \cite{bauer2023quantum,banuls2020review,di2024quantum,humble2022snowmass}.
 
While quantum computing represents an enormous scientific opportunity, realizing its potential requires both experimental and theoretical advances. Progress has been made in experimental demonstrations of lattice gauge theories in one and two spatial dimensions (1D and 2D) \cite{martinez_real-time_2016, rahman2022, martinez_compiling_2016,atas2021,paulson2021simulating,meth2023simulating,atas2021,atas2023,yang_observation_2020,farrell2024scalable,kavaki2024square,nguyen_digital_2022,farrell2024quantum,davoudi2024scattering,klco_su2_2020,ciavarella_trailhead_2021,zhou2022thermalization,ciavarella2024string,su2023observation,de2010simulating,mueller2023quantum,turro2024qutrit}. 
These lower-dimensional models can capture interesting physics with reduced resource requirements and serve as a pathway to quantum simulations in higher dimensions. However, even in 1D, quantum simulations of phase diagrams face significant hurdles that suggest the need for new approaches. The first one is the preparation of mixed states on quantum computers \cite{verdon2019quantum,wu_variational_2019,wang_variational_2021,sagastizabal_variational_2021,foldager_noise-assisted_2022,fromm2023simulating,selisko2023extending,guo2021thermal,xie2022variational,davoudi2023towards,zhu2020generation,consiglio2023variational,zhou2022thermalization}. This is difficult since it requires non-unitary evolution or thermal sampling of multiple eigenstates.
The second one is ensuring the color-neutrality constraints imposed by the boundary conditions and the gauge symmetry of the model. 

In this article, we overcome these challenges by introducing (i) motional ancillae to efficiently create thermal probability distributions on a trapped ion device, and (ii)  charge-singlet measurements that use group-theoretical projector techniques.

\begin{figure*}[thb!] 
    \centering
\includegraphics[width=1.0\textwidth]{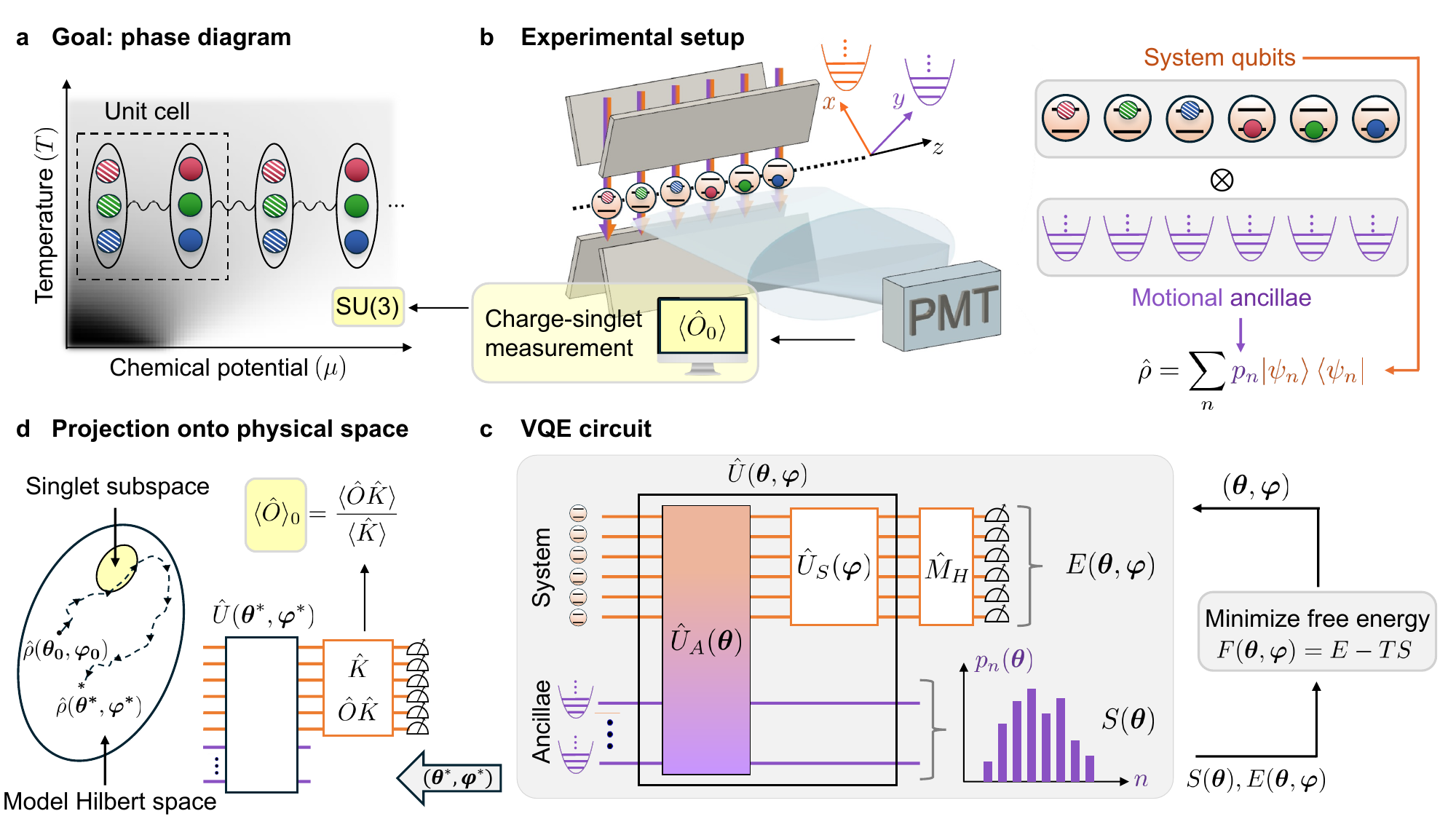} 
    \caption{\textbf{Particle physics phase diagram on a quantum computer.} (a) We study the SU$(2)$ and SU$(3)$ phase diagram on a 1D lattice by preparing thermal states at finite chemical potential $\mu$. A unit cell consists of an antimatter site (striped circles) and a matter site (solid circles), connected by a gauge field (wiggly line). (b) In our experiment, each ion acts as a qubit, encoding quark color components in its internal states. For $N$ ions, $N$ motional modes in the $y$-direction serve as an ancilla register (purple), and $N$ motional modes in the $x$-direction mediate entangling gates between qubits (orange). Qubit and motional operations are driven by a set of addressed laser beams, and the qubit states are measured by collecting fluorescence on a photo-multiplier tube (PMT) array. 
    (c) A parametrized circuit $\hat{U}_A(\boldsymbol{\theta})$ prepares a probability distribution $p_{n}(\boldsymbol{\theta})$, used to calculate the entropy $S(\boldsymbol{\theta})$ of the thermal state.  
    The resulting distribution of initial states in the system register is subject to a second parametrized circuit $\hat{U}_S(\boldsymbol{\varphi})$. The energy $E(\boldsymbol{\theta}, \boldsymbol{\varphi})=\langle \hat{H}\rangle$, is measured by suitably rotating the measurement basis using additional unitaries $\hat{M}_H$. Using the measured energy and entropy values, the free energy is calculated and classically minimised to find optimal parameters $(\boldsymbol{\theta}^*,\boldsymbol{\varphi}^*)$ for a given temperature and chemical potential. (d)  Our unconstrained variational search (dashed path) explores the model Hilbert space (large oval). A projection method retrieves the expectation value $\langle \hat{O}\rangle_{0}$ of an observable $\hat{O}$ within the charge-singlet subspace as the ratio of $\langle\hat{O}\hat{K}\rangle$ and $\langle\hat{K}\rangle$, where $\hat{K}$ is a projection operator specific to the underlying gauge group.}
    \label{fig:protocol}
\end{figure*}

To address the first challenge, we leverage a subset of the 3$N$ vibrational modes that are naturally available in a system of $N$ trapped-ion qubits as motional ancillae~\cite{leibfried2003quantum}. By employing these motional modes beyond their typical role as intermediaries of entangling qubit operations, we add an independent ancilla register without increasing the system size. Our approach paves the way for the applications of motional modes in the study of thermal states in quantum many-body systems.

The second challenge requires enforcing the non-Abelian gauge symmetry constraints \cite{kogut_hamiltonian_1975}.  In the absence of background charges, the thermal state should be a probabilistic mixture of color-neutral states, i.e., states with zero global color charge. Instead of requiring the thermal state to respect these symmetry constraints, we incorporate them into the measurement process. This strategy uses a group-theoretical projection technique \cite{elze1986quantum,leyaouanc1989,mclerran1985thermodynamics} that projects a state onto the color-neutral or singlet subspace and enhances our protocol's flexibility, facilitating an effective implementation of the color-neutrality condition.

In combination, the above methods allow us to perform the first experimental study of the phase diagram of SU$(2)$ and SU$(3)$ non-Abelian gauge theories with dynamical matter on a quantum computer.

\section*{Thermal states in gauge theories}
Gauge theories are the backbone of the Standard Model of particle physics, with QCD using the SU$(3)$ gauge group to describe the interaction between quarks (fermions) and gluons (gauge bosons). We study the gauge groups SU$(2)$ and SU$(3)$---with two and three colors, respectively---in
1D with open boundary conditions, with dynamical fermionic matter at nonzero temperature $T$, and chemical potential $\mu$.

We use here the Kogut-Susskind Hamiltonian approach~\cite{kogut_hamiltonian_1975} to lattice gauge theory (LGT), that describes the interaction between (fermionic) matter fields  and (bosonic) gauge fields defined on the vertices of the lattice and on the links between vertices, respectively (see Fig.~\ref{fig:protocol}a).  In natural units $(\hbar = c = k_B = 1)$, the Hamiltonian consists of the following terms 
\begin{equation}
    \hat{H}=\hat{H}_{kin}+ am \hat{H}_{mass}+\frac{1}{2x}\hat{H}_{elec}-a\mu \hat{H}_{chem} \;, \label{eq:Hamiltonian-symbolic}
\end{equation}
where the first term is the kinetic energy and describes how matter fields interact with gauge fields as they move between lattice sites. The second term encodes the mass contribution of the matter fields, where $m$ denotes the bare fermion mass and $a$ is the lattice spacing. The third term is the color electric field energy contribution, where $x= 1/(ga)^{2}$ is related to the gauge-matter coupling strength $g$. Finally, the last term describes the matter-antimatter imbalance in the system and accounts for nonzero chemical potential $\mu$. Throughout the rest of this work, we will adopt the conventional lattice units where $a=1$, making the temperature and chemical potential dimensionless.  Explicit form of this Hamiltonian in terms of fermionic and gauge fields are given in the Supplementary information ~\ref{sec:su2_general} and \ref{sec:su3_general}).

The theory’s non-Abelian local gauge symmetry leads to a set of non-commuting conserved charges that give rise to constraints known as Gauss laws (see Supplementary Information~\ref{sec:su2_general} and \ref{sec:su3_general}).  Physical states satisfy these Gauss laws and are referred to as the gauge-invariant states. We use the Gauss constraints to eliminate the gauge fields  \cite{martinez_real-time_2016,sala_variational_2018,atas2021}, resulting in a purely fermionic Hamiltonian that is then mapped onto qubits by a Jordan-Wigner (JW) transformation (see Supplementary Information~\ref{sec:su2_general} and \ref{sec:su3_general}).  Elimination of the gauge fields using JW transformations results in long-range four-body interactions for SU(2) and six-body interactions for SU(3). In this work, we focus on the phase diagram of a single unit cell, where such long-range interactions are absent. Nonetheless, it is possible to implement these interactions on a trapped-ion platform due to the intrinsic all-to-all connectivity between the ions~\cite{martinez_real-time_2016,nguyen_digital_2022}.

 After eliminating the local gauge degrees of freedom, we restrict our analysis to the sector with zero global color charge, as required by the boundary condition that assumes the absence of background charges. This choice is motivated by the physical observation that hadrons with non-zero color charge are not observed in nature. States that satisfy this zero global charge condition are referred to as charge-singlet states (see Fig.~\ref{fig:protocol}d). Our goal is for the thermal state to be a probabilistic mixture of such charge-singlet states at a given temperature.

To study the phase diagram (see Fig.~\ref{fig:protocol}a), we prepare the Gibbs thermal states at temperature $\beta=T^{-1}$, 
\begin{equation}
    \hat{\rho}_{G}=\frac{e^{-\beta \hat{H}}}{Z}=\sum_n p_n \ket{E_n}\bra{E_n}, \;\; Z=\text{Tr} \, \left( e^{-\beta \hat
    {H}}\right). \label{eq:gibbs_thermal}
\end{equation}
Here, $Z$ is the partition function and ensures proper normalisation. The thermal density matrix  $\hat{\rho}_G$  is a probabilistic mixture of pure states as can be seen from the expansion in the eigenstate basis $\ket{E_{n}}$ of the Hamiltonian $\hat{H}$,
% \begin{equation}
%     \hat{\rho}_G = \sum_n p_n \ket{E_n}\bra{E_n}, \label{eq:prob_mixture}
% \end{equation}
where $p_n = e^{-\beta E_n}/Z$ are the Boltzmann weights and $E_{n}$ are the eigenvalues of the Hamiltonian. The Gibbs state minimises the free energy $F = E - TS$, where $E = \text{Tr}(\hat{\rho}_{G}\hat{H})$ is the internal energy of the system and $S = -\text{Tr}(\hat{\rho}_G \ln \hat{\rho}_G)$ is the entropy.

Importantly, to construct the  charge-singlet thermal state from $\hat{\rho}_{G}$, the trace in the partition function and the eigenstates $\ket{E_{n}}$ must be restricted to the  singlet subspace.  This requirement adds significant complexity to the process of preparing thermal states for LGTs.

A natural choice of order parameter for exploring the phase diagram is the chiral condensate $\hat{\chi}$, which is proportional to the mass term 
$\hat{H}_{mass}$. The explicit form of this operator is provided in Methods  \ref{app:su2_chiral_cond} and \ref{app:su3_chiral_cond} for the SU(2) and SU(3) cases, respectively. In (3+1)-D QCD, the chiral condensate serves as the order parameter for the $T-\mu$ phase diagram. A change from a negative chiral condensate value (corresponding to condensate formation) to zero chiral condensate (corresponding to quark-gluon plasma formation) as $\mu$ is varied in (3+1)-D QCD signifies a phase transition from chiral symmetry broken phase to chiral symmetry restored phase. However, the precise location and nature of the critical points in the QCD phase diagram at finite density remains unknown due to the sign problem. Motivated by this open question in fundamental physics, we adopt $\hat{\chi}$ as the order parameter in our (1+1)-D lattice QCD simulation. Our goal is to observe the same transition from a negative to zero chiral condensate, which in our context reflects a change in the dominant eigenstate contributions to the thermal density matrix—an interpretation that we elaborate upon in the discussion of the experimental results below.

\section*{Thermal state preparation with motional ancillae}\label{sec:thermal_state}

We describe here our protocol for preparing thermal states in gauge theories on a quantum computer and present the two key components that enable it: the introduction of motional ancillae and  charge-singlet measurements.

\textit{Experimental setup:} 
A chain of $^{171}$Yb$^+$ ions is held in a linear Paul trap (Fig.~\ref{fig:protocol}b). Each ion provides a qubit in its internal degrees of freedom, encoded in the hyperfine-split electronic ground level, with $\ket{0} = \ket{^2S_{1/2}, F=0, m_F=0}$ and $\ket{1} = \ket{^2S_{1/2}, F=1, m_F=0}$. An array of $N$ ions in the trap also provides three sets of $N$ orthogonal harmonic motional modes, one in each spatial direction. 
We utilise the modes in the radial $x$-direction as intermediaries to realise standard entangling gate operations between qubits with a variant of the M\o lmer-S\o rensen (MS) gate scheme (see Methods~\ref{app:exp_platform} for details). The modes in the $y$-direction are typically unused, but here we leverage them as an independent ancilla register for preparing the probabilities in the Gibbs ensemble.

\textit{Protocol:} To prepare the Gibbs state, we use a variational quantum eigensolver (VQE) \cite{mcclean_theory_2016} to find the state that minimises the free energy \cite{consiglio2023variational}.
Our parametrized VQE circuit (Fig.~\ref{fig:protocol}c) consists of two parts.
The first unitary circuit $\hat{U}_A(\boldsymbol{\theta})$ couples the motional ancillae with the system register.
When the ancilla register is traced out, it creates a tunable probability distribution $\tilde{p}_{j}(\boldsymbol{\theta})$. For a general probability distribution, the unitary $\hat{U}_A(\rm\boldsymbol{\theta})$ 
contains a sequence of parametrized single-qubit and entangling gates acting on the motional ancillae, followed by gates entangling the motional ancillae with the system qubits.  For the system considered in this work, it is sufficient to create the pairwise qubit-motion entangled state $\cos(\theta_i/2)\ket{0,0}+ \sin(\theta_i/2)\ket{1,1}$, where the computational basis is denoted by $\ket{\text{spin, motion}}$. In a trapped-ion system, this qubit-motion entangled state is created using a partial sideband rotation (see Methods~\ref{app_thermal}) with a laser drive detuned by the motional frequency. By tuning $\theta_i$ for each qubit-motional mode pair, and by tracing out the motional ancillae, the entire system register is prepared in a mixture of bit strings $\ket{j} \equiv \ket{j_1j_2\cdots j_N}$, with a probability $\tilde{p}_{j}(\boldsymbol{\theta})=\prod_{i=1}^{N}\tilde{p}_{j_i}(\theta_i)$. A more complex set of unitary operations on the motional ancillae ~\cite{hou2024coherent} will lead to more general probabilities.

A second  unitary circuit $\hat{U}_S(\boldsymbol{\varphi})$ is applied on the qubit register to create the eigenstates of the density matrix. In combination with the motional ancillae, this produces the following variational ansatz for the density matrix
\begin{equation}
    \hat{\rho}(\boldsymbol{\theta},\boldsymbol{\varphi})=\sum_{j} \tilde{p}_{j}(\boldsymbol{\theta})\;\hat{U}_{S}(\boldsymbol{\varphi})\ket{j}\bra{j}\hat{U}^{\dagger}_{S}(\boldsymbol{\varphi}).\label{density_mat_ansatz}
\end{equation}
The variational parameters $(\boldsymbol{\theta},\boldsymbol{\varphi})$ are then updated through a feedback loop between a classical optimiser and the ion trap to minimise the cost function $F[ \hat{\rho}(\boldsymbol{\theta},\boldsymbol{\varphi})]=\mathrm{Tr}( \hat{\rho}(\boldsymbol{\theta},\boldsymbol{\varphi}) \hat{H})- T\langle\hat{S}\rangle (\boldsymbol{\theta})$. The average of the Hamiltonian in the cost function is measured on the system register. For a general construction of the unitary $\hat{U}_A(\boldsymbol{\theta})$, the entropy can be obtained by measuring the ancilla register $\langle\hat{S}\rangle (\boldsymbol{\theta}) = -\sum_j \tilde{p}_j(\boldsymbol{\theta}) \log \tilde{p}_j(\boldsymbol{\theta})$.
For the design of $\hat{U}_{A}$ in our experiment, the entropy can be calculated analytically (see Methods~\ref{app:su2_vqe_circ} for more details) from the partial sideband rotation angles $\theta_i$ in the ancilla using
\begin{align}
\notag S(\boldsymbol{\theta})=-\sum_{i} & \left[\cos^2{(\theta_i/2)}\log(\cos^2{(\theta_i/2)}) \right. \\ &+\left.\sin^2{(\theta_i/2)}\log(\sin^2{(\theta_i/2)})\right] \;.\label{eq:entropy_analytical}
\end{align}
Consequently, measurements of the motional ancillae are not required in our protocol. Furthermore, due to the structure of our ancilla circuit, the probability distribution is insensitive to the relative phases of the motional modes. This makes our experiment resilient to motional decoherence and imperfect cooling, bypassing the typical hurdles in using motional modes for computational tasks~\cite{hou2024coherent}.

At the end of the variational search, using the optimal parameters $(\boldsymbol{\theta}^*,\boldsymbol{\varphi^*})$, we prepare the Gibbs state in Eq.~(\ref{eq:gibbs_thermal}) on our trapped-ion device.  If the VQE converges successfully, the probabilities $\tilde{p}_j(\boldsymbol{\theta^*})$  will match the Boltzmann weights $p_n$ in Eq.~(\ref{eq:gibbs_thermal}) and the states $\hat{U}_{S}(\boldsymbol{\varphi^*})\ket{j}$ will approximate the basis vectors $\ket{\psi_n}$ of the thermal density matrix. For non-degenerate eigenvalues $p_n$, these $\ket{\psi_n}$s are energy eigenstates $\ket{E_n}$ and otherwise they form orthonormal linear combinations of the energy eigenstates within the degenerate subspace. Importantly, the basis states $\ket{\psi_n}$ are obtained through variational optimization of the cost function without requiring any prior knowledge of the energy eigenstates and the underlying spectrum.

So far, we have not incorporated the  global charge constraints in our protocol. Hence, the eigenstates of this density matrix are not restricted to the charge-singlet subspace. To distinguish the prepared thermal state from the density matrix restricted to the  singlet subspace, we refer to it as the unconstrained density matrix.

\textit{Charge-singlet measurements:} To implement the  global charge constraint, we modify the measurement of physical observables to yield the same expectation values as those obtained from the thermal state restricted to the  singlet subspace.
For a given physical observable $\hat{O}$, we  use a group-theoretical projection method \cite{elze1986quantum,leyaouanc1989} to define the   observable expectation value $\langle\hat{O}\rangle_{0}$ on the singlet subspace as
 \begin{equation}
     \langle \hat{O}\rangle_{0}= \frac{\langle \hat{O}\hat{K}\rangle}{\langle \hat{K}\rangle}\;. \label{eq:projection_formula}
 \end{equation}
Here, the averages on the right hand side are measured with respect to the unconstrained density matrix $\hat{\rho}(\boldsymbol{\theta}^*,\boldsymbol{\varphi}^*)$   prepared on the device at the end of the variational optimisation and $\hat{K}$ is a projection operator specific to the group under consideration. 
In  Methods~\ref{app:projection_operator}, we explain in more detail how the projection operator $\hat{K}$ can be calculated explicitly for SU$(2)$ and SU$(3)$ groups. 
In our case, the order parameter $\langle\hat{\chi}\rangle_{0}$ is evaluated using Eq.~(\ref{eq:projection_formula}). Both $\hat{\chi}\hat{K}$ and $\hat{K}$ are diagonal operators (see Methods~\ref{app:su2_chiral_cond}, \ref{app:su3_chiral_cond}, and Supplementary Information \ref{app:projector_integration}), and can thus be measured using $\hat{\sigma}^{z}-$basis measurements only.  Eq.~(\ref{eq:projection_formula}) can be used for determining the expectation value of non-diagonal observables (e.g., the Hamiltonian) from the unconstrained density matrix but would require non-diagonal Pauli measurements to be performed on the system due to $\hat{O}\hat{K}$ consisting of non-diagonal Pauli strings.

By determining $\langle\hat{\chi}\rangle_{0}$ using the charge-singlet measurement technique and leveraging motional ancillae, we can now investigate the phase diagram on a trapped-ion device.

\section*{SU$(2)$ and SU$(3)$ phase diagram on an ion-trap quantum computer}
\begin{figure*}[!ht] %
    \centering
    \includegraphics[width=1.0\textwidth]{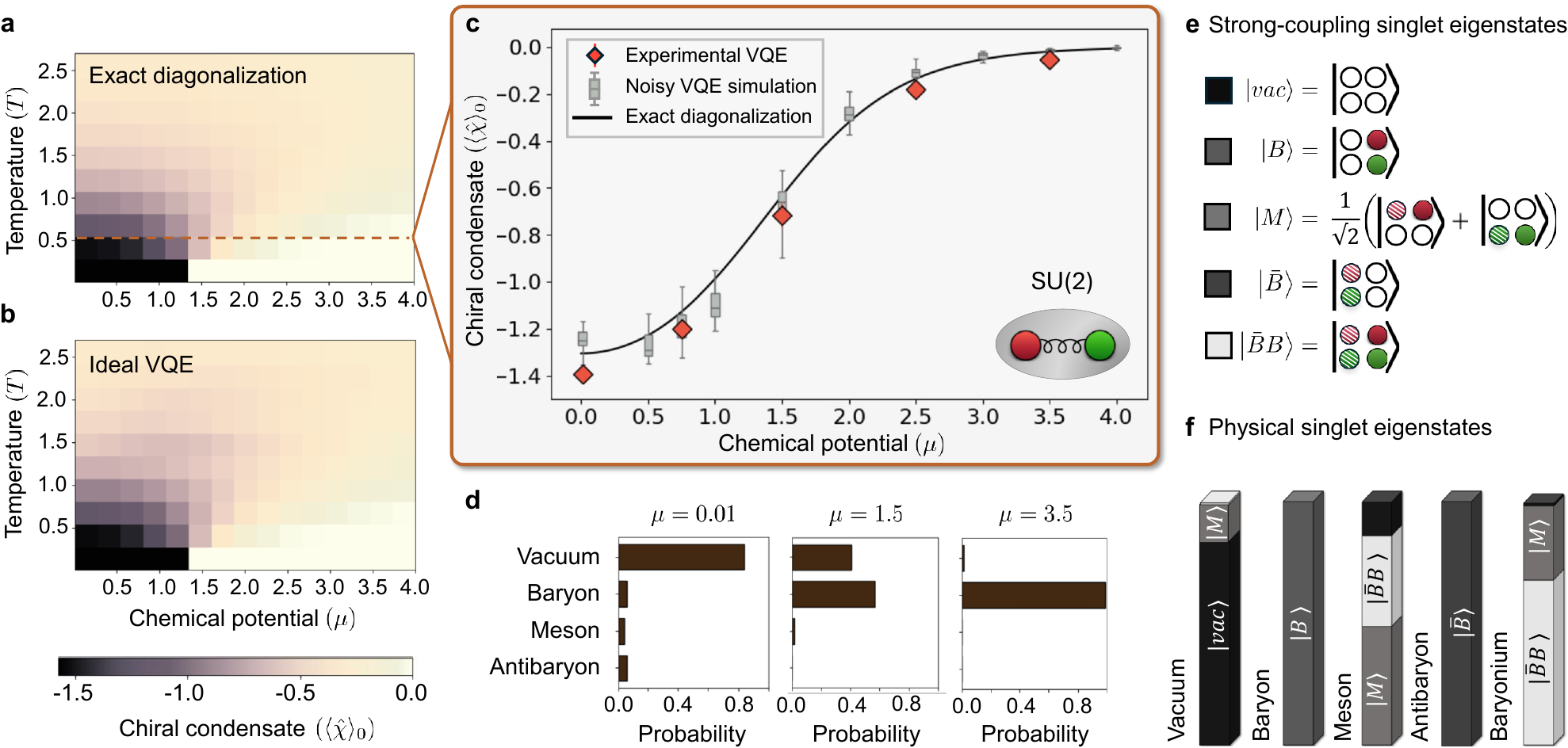} 
    \caption{\textbf{SU(2) thermal states for a unit cell with trapped ions.} (a) Exact diagonalization (ED) results for the SU(2) unit cell for $x=1$ and $m=0.5$. The order parameter $\langle\hat{\chi}\rangle_{0}$ (chiral condensate) takes large negative values in the low $T$ and $\mu$ limit. Chiral symmetry $\langle\hat{\chi}\rangle_{0}$=0 is restored at high $\mu$ and $T\rightarrow \infty$. (b) Classical simulation results for our variational protocol (Fig.~\ref{fig:protocol}) for the noise-free case. (c) Experimental data for $T=0.5$ (dashed line in panel (a)). Our motional ancillae based protocol uses up to 230 cost function evaluations per point, determining the chiral condensate for five distinct chemical potential values. The experimental VQE results (red diamonds) are in good agreement with both the ED (black curve) and noisy simulation results (grey boxes). The grey boxes show the spread of mean chiral condensate values from twenty noisy VQE runs (represented by the error bar with the box denoting the inter-quartile range) for each chemical potential, highlighting the protocol’s high success rate. (d) Composition of the charge-singlet thermal state at varying chemical potentials. The mixtures of SU(2) physical eigenstates show the transition from a vacuum-dominated to a baryon-dominated phase. Panel (f) shows the composition of the physical eigenstates in terms of the strong coupling ($x \ll 1$) eigenstates (panel (e)). The heights of the various bar-segments represent the contributions of the strong-coupling states.
    }
    \label{fig:targetplot}
\end{figure*}

We implement our protocol for the LGT unit cell (Fig.~\ref{fig:protocol}a), which hosts red and green quarks and anti-quarks for SU$(2)$, and additional blue quarks and anti-quarks for SU$(3)$. The Hamiltonians for each case are given explicitly in Methods~\ref{app:su2_unit_cell} and \ref{app:su3_unit_cell}. For the experiment, we choose the Hamiltonian parameters $x=1$ and $m=0.5$ in Eq.~(\ref{eq:Hamiltonian-symbolic}) for both models, placing the system in the intermediate coupling strength regime, where neither the electric field nor the mass term dominates.

\textit{Experimental realisation:} 
The variational circuit $\hat{U}_S(\boldsymbol{\varphi})$ in Fig.~\ref{fig:protocol}c consists of gates that implement different terms in the target Hamiltonian in Eq.~(\ref{eq:Hamiltonian-symbolic}), as shown in  Methods~\ref{app:su2_vqe_circ} and \ref{app:su3_vqe_circ}. The SU$(2)$ and SU$(3)$ LGT circuit contain three- and four-qubit gates, respectively, which can be expressed in terms of the native two-qubit MS gates (see Supplementary Information~\ref{app:circ_decomp}). This circuit, combined with the parametrized partial sideband rotations that couple the motional ancillae with the system register, completes our ansatz. The VQE parameters (10 for SU$(2)$ and 21 for SU$(3)$) are then classically optimised using a Bayesian direct search algorithm \cite{acerbi2017practical}, where the cost function is evaluated on the quantum computer.

The VQE cost function $F = E-TS$ consists of two components: energy $E$ and entropy $S$. The entropy term is computed analytically using Eq.~(\ref{eq:entropy_analytical}) based on the gate angles of the partial sideband rotations. The energy is measured on the system register using the Pauli-decomposition of the Hamiltonian. In both SU(2) and SU(3) LGT, the Hamiltonian decomposes into two families of commuting Pauli strings: one comprising the diagonal terms and the other comprising the non-diagonal terms (Methods~\ref{app:su2_unit_cell} and \ref{app:su3_unit_cell}). 
The diagonal terms can be measured directly in the $\hat{\sigma}^{z}-$basis. We design a measurement circuit $\hat{M}_H$ using the formalism developed in \cite{gokhale2019minimizing} to evaluate the expectation value of all the non-diagonal Pauli strings simultaneously. This reduces the number of required measurements at the cost of adding a small number of entangling gates (see  Methods~\ref{app:su2_vqe_circ} and \ref{app:su3_vqe_circ}). Each $\hat{\sigma}^{z}-$basis measurement is repeated 2000 and 3000 times for SU$(2)$ and SU$(3)$, respectively. As the chemical potential $\mu$ varies, the structure of the VQE ansatz remains unchanged; however, the optimised parameters $(\boldsymbol{\theta^*},\boldsymbol{\varphi^*})$ change accordingly, resulting in a different composition of the optimised density matrix for each value of $\mu$.

\begin{figure*}[!ht] 
    \centering
    \includegraphics[width=1.0\textwidth]{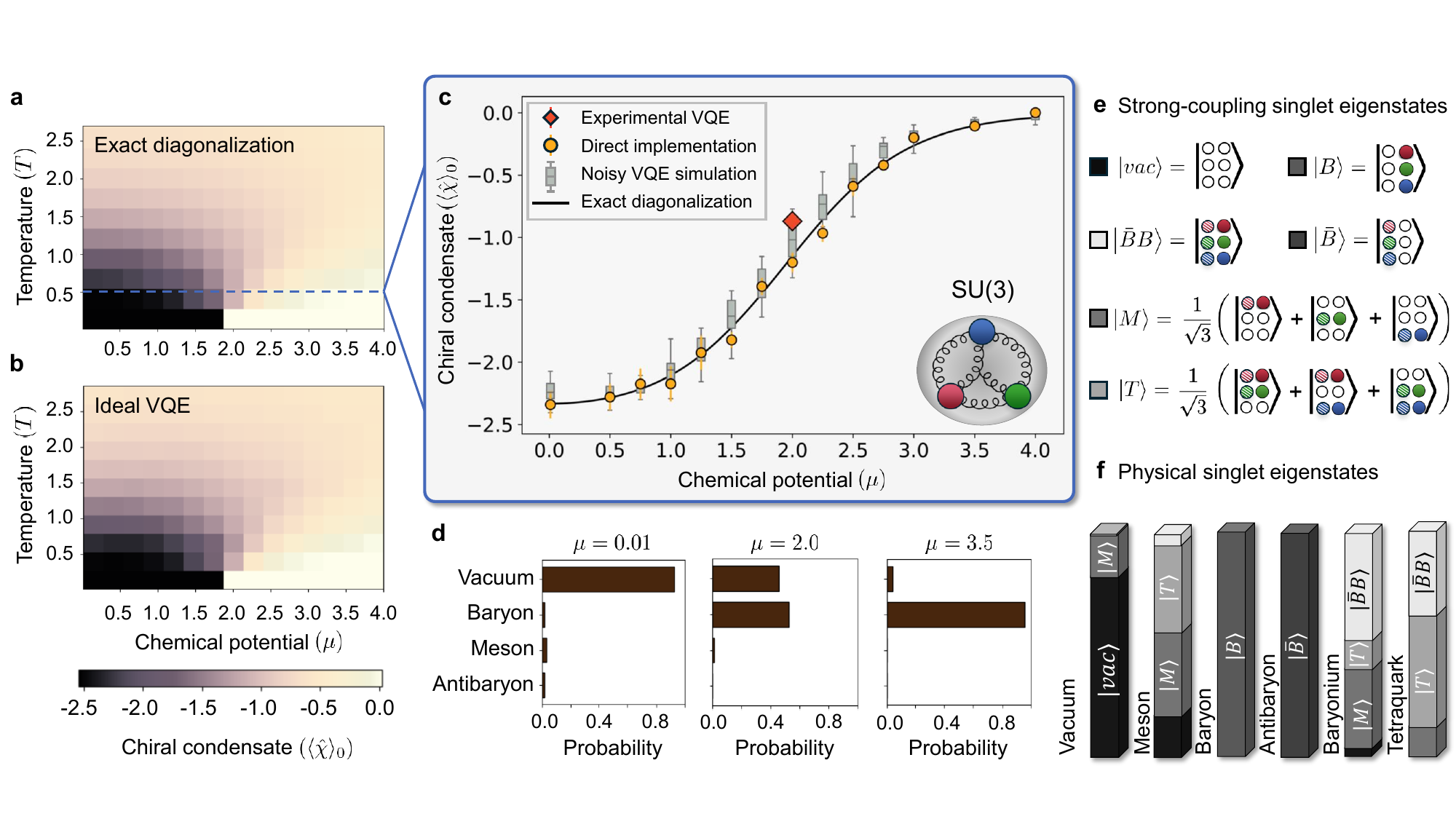} 
    \caption{\textbf{SU(3) thermal states for a unit cell with trapped ions}. (a) Chiral condensate for a unit cell obtained from exact diagonalization (ED) for $x=1.0, m = 0.5$. The phase diagram is qualitatively similar to Fig. \ref{fig:targetplot}a, but differs quantitatively, with the transition point at zero temperature occurring at a distinct $\mu$-value compared to SU(2). (b) Classical simulation results for our variational protocol (Fig. \ref{fig:protocol}) in the noiseless case. (c) The VQE experiment is run for $\mu = 2$ close to the phase transition, allowing up to 350 cost function evaluations. The experimental result matches well with the noisy VQE simulation, showing the effectiveness of the ansatz in preparing the thermal state near the transition. Additionally, the VQE circuit is run using the optimised ideal VQE parameters for $T = 0.5$ for a range of $\mu$ values, confirming our noise model. The spread of the noisy VQE simulation collected over twenty trials (represented by the error bar with the box denoting the interquartile range) highlights the reliability of our protocol. (d) Boltzmann weights of eigenstates of the Hamiltonian in the charge-singlet thermal state are shown at three different chemical potentials, highlighting the transition from vacuum-dominated density matrix to baryon-dominated density matrix. Panels (e) and (f) show the strong coupling ($x\ll 1$) and physical eigenstates. Due to the presence of three colors, the unit cell allows for more gauge-invariant states than the SU(2) model in Fig.~\ref{fig:targetplot}, which did not include the tetraquark state.}
    \label{fig:targetplot2}
\end{figure*}

To validate the performance of our VQE scheme, we run ideal simulations of our protocol for different temperature and chemical potential values with the results shown in Figs.~\ref{fig:targetplot}b and \ref{fig:targetplot2}b. For the experiment, we fix the temperature at $T=0.5$, where the chiral condensate shows a phase transition with $\mu$. In this intermediate temperature regime, conventional action-based methods face difficulties in probing the phase diagram for larger lattice sizes.

\textit{SU(2) thermal states:} The full variational protocol is executed on the trapped-ion system for five different values of the chemical potential, shown in
Fig.~\ref{fig:targetplot}c. The results show excellent agreement with exact diagonalization. At very low chemical potential, the chiral condensate is strongly negative, gradually rising towards zero as the chemical potential is increased.  This behavior resembles the phase transition from chiral symmetry broken phase to chiral symmetry restored phase in (3+1)-D QCD. The large negative values of the chiral condensate indicates a thermal mixture dominated by the physical vacuum state, whereas a zero value of chiral condensate corresponds to a mixture dominated by the baryon state. This can be explained from the expression of the Boltzmann weights $e^{-\beta E_n} = e^{-\beta(\tilde{E}_n-\mu B)}$, where $\tilde{E}_n$ are the energy eigenvalues corresponding to the eigenstates $\ket{E_n}$ at zero chemical potential and $B$ is the baryon number (expectation value of $\hat{H}_{chem}$). At low chemical potential, the energy contribution $\tilde{E}_n$ dominates in the Boltzmann weight, leading to a thermal mixture where the lowest energy eigenstate, i.e., the physical vacuum is energetically favored. On the contrary, at high $\mu$, the baryon contribution $\mu B$ dominates over the energy $\tilde{E}_n$, leading to a baryon-dominated density matrix. 

% This behavior indicates that chiral symmetry is broken at low values of $\mu$ but is restored at higher $\mu$, where the presence of baryons becomes energetically favored.  

Fig.~\ref{fig:targetplot}d shows the probabilities (the Boltzmann weights) of the  physical singlet eigenstates of the Hamiltonian that appear in the thermal state mixture.
This phase transition from vacuum to baryon-dominated density matrix is effectively captured by our experiment. Additionally, we employ a noise model that simulates the effects of the dominant noise sources (see Methods~\ref{app:noise_model}) in our experimental device and conduct multiple noisy VQE trials to assess our ansatz's performance. The distribution of chiral condensate values for different $\mu$ in Fig.~\ref{fig:targetplot}c (represented by the grey boxes) underscores the reliability of the protocol.

The physical color-neutral eigenstates in Fig.~\ref{fig:targetplot}d are a linear superposition of the strong-coupling color-neutral eigenstates (Fig.~2e), which form a convenient basis for the  subspace. For example, the physical vacuum state is different from the strong-coupling state $\ket{vac}$, which corresponds to an all-empty lattice configuration. Fig.~\ref{fig:targetplot}f shows the composition of the physical eigenstates in terms of the strong coupling basis. The contributions of these basis states to the physical eigenstates are governed by the choice of the Hamiltonian parameters $x$ and $m$, but do not depend on $T$ and $\mu$. For fixed $x$ and $m$ values, the temperature and chemical potential determine the Boltzmann weight of each physical eigenstate in the mixture, which determines the shape of the phase curve in Fig.~\ref{fig:targetplot}c. In particular, as temperature increases, the transition from the vacuum-dominated to baryon-dominated thermal density matrix becomes smoother, eventually disappearing at high $T$. This produces a localized region of negative chiral condensate values in the lower left corner of the phase diagram in Fig.~\ref{fig:targetplot}a.

\textit{SU(3) thermal states:} Including an additional color (blue) allows for richer physics compared to SU$(2)$.
In particular, an SU$(3)$ baryon is composed of three colored quarks and exhibits true fermionic behavior, unlike the SU$(2)$ model, where a baryon consists of only two quarks and follows bosonic statistics. As a result, the SU$(3)$ model encounters the sign problem, while the SU$(2)$ model does not. This makes SU$(3)$ LGT an ideal candidate for leveraging quantum computers. Additionally, the enlarged Hilbert space for SU$(3)$ allows more degrees of freedom in constructing charge-singlet states on a lattice.

The increased non-locality of the interactions present in the Hamiltonian as well as the larger size of the Hilbert space makes the experimental realisation of the VQE more demanding.  
Again, we chose intermediate parameter values, in this case $\mu=2$ and $T=0.5$, at the phase transition from a vacuum-dominated phase to a baryon-dominated phase, see Fig.~\ref{fig:targetplot2}d. Our experimental VQE successfully prepares the thermal mixture, which combined with charge-singlet measurements for the chiral condensate agrees with the exact diagonalization value of $\langle\hat{\chi}\rangle_{0}$ (Fig.~\ref{fig:targetplot2}c).

In SU$(3)$, constructing the physical vacuum is more involved than SU$(2)$ (Fig.~\ref{fig:targetplot2}f) due to the presence of an additional strong-coupling singlet eigenstate in the superposition, which we call the tetraquark state~\cite{atas2023}. Absent in the unit cell of SU$(2)$, the tetraquark state consists of a pair of quarks and a pair of antiquarks (Fig.~\ref{fig:targetplot2}e). 
Our successful VQE optimisation for the phase transition point, where contributions from the physical vacuum and baryon state become of similar magnitude, demonstrates the capability to capture both vacuum- and baryon-dominated phases effectively. Demonstrating the successful VQE performance for the transition point provides strong evidence that the approach is expected to work at other values of the chemical potential as well.
It is thus a significant first step towards simulating the whole phase diagram experimentally for larger lattice sizes.

We use the same device-aware noise model as before for SU$(3)$, and run multiple independent numerical trials of the noisy VQE for different values of the chemical potential. Figure \ref{fig:targetplot2}c shows that the noisy VQE successfully traces the path from a chiral symmetry-broken phase to a symmetry-restored phase. The larger spread of the error bars from the noisy VQE indicates that although it is more error-prone compared to SU$(2)$, the VQE still performs well in estimating the chiral condensate value. The larger fluctuations in optimization are a consequence of increased parameter space, register size, and circuit depth. Additionally, we observe that the fluctuations are most pronounced for intermediate values of $\mu$, near the phase transition point. This aligns with our expectation that this region is the most challenging to capture reliably in experiments, motivating our choice of $\mu$ for the experiment.
Furthermore, we use the optimised parameters obtained from our noisy VQE simulations to directly prepare the unconstrained thermal state on our trapped-ion experiment and evaluate the chiral condensate. Our measurements yield excellent agreement with the simulated VQE and exact diagonalization results, validating our noise model.

%\section*{Prospects for other gauge theories and many-body physics}
\section*{Discussion}
Quantum simulations of particle physics have so far mostly focused on pure states at $T=0$  \cite{banuls_simulating_2020,bauer2023quantum, di2024quantum}. However, to describe nature, it is crucial to understand states of matter at finite temperature. Our work opens the door to resource-efficient quantum simulations of thermal states in gauge theories.

Our charge-singlet measurement technique is broadly applicable for different gauge theories and can be easily extended to studying dynamics using e.g. Trotter time evolutions.  The projection-based technique can be extended to two or three spatial dimensions, when it is no longer possible to integrate out all gauge degrees of freedom. In these higher-dimensional settings, the charge-singlet projection method can be generalised to enforce both the local Gauss law at each vertex and the global charge constraints, offering an alternative to explicitly imposing gauge invariance. This generalised projection is also applied during the measurement of observables.

Beyond quantum computing, the projector technique used here is equally valuable for classical Hamiltonian-based computations, such as tensor network state calculations. For both quantum- and classical simulations, charge-singlet measurements could be employed to explore thermodynamic quantities like entropy and work within charge-singlet subspaces, offering exciting links to quantum thermodynamics in gauge theories \cite{davoudi2024quantum,majidy2023noncommuting}. 

%Our gauge-invariant measurement technique is broadly applicable for different gauge theories and can be easily extended to study dynamics of thermal states. The projector technique used here is equally valuable for classical Hamiltonian-based computations, such as tensor network state calculations. Notably, gauge-invariant measurements could be employed to explore thermodynamic quantities like entropy and work within gauge-invariant subspaces, offering exciting links to quantum thermodynamics in gauge theories \cite{davoudi2024quantum,majidy2023noncommuting}.

An important next step toward simulating particle physics is the extension to two and ultimately three dimensions. One advantage of using quantum computers for LGT simulations is that the computational framework can be generalised without significant theoretical roadblock. While the concepts from previous works are transferable (see also~\cite{kan_investigating_2021,paulson2021simulating,haase_resource_2021,meth2023simulating,zohar_quantum_2015,zohar2022quantum}), scaling up the lattice size will be crucial, increasing the need for resource-efficient methods. Moreover, incorporating quark flavors and designing protocols to connect future quantum simulations with observables in particle physics are also interesting topics for future work.

%The next step toward simulating particle physics is extending the models to two and ultimately three dimensions. While the concepts from previous works are transferable (see also~\cite{kan_investigating_2021,paulson2021simulating,haase_resource_2021,meth2023simulating}), scaling up the lattice size will be crucial, increasing the need for resource-efficient methods. Incorporating quark flavors and designing protocols to connect future quantum simulations with observables in particle physics are also interesting topics for future work.

Our resource-efficient motional ancillae approach leverages otherwise unused degrees of freedom and can be further developed into a fully functional qubit register. utilising motional modes in ions for certain special applications is already well underway \cite{Zahringer2010,Toyoda2015,Zhang2018,Chen_2021,Chen2023,Whitlow2023,davoudi2021toward,navickas2024experimental,Valahu2024}, with ongoing efforts realising motional qubits capable of readout, performing general unitary \cite{hou2024coherent} and state-dependent operations \cite{Vasquez_arxiv2024}.
Similar to the all-to-all connectivity available with entangling gates, these ancillary states can couple with nearly any qubit in the system register. This capability holds the potential to create arbitrary probability distributions for general thermal states—using, e.g., circuits inspired by autoregressive models \cite{liu2021solving}. Our proof-of-concept can also be adapted to bosonic modes that remain otherwise idle in other quantum systems, such as cavity quantum electrodynamics and superconducting circuit platforms.

%Our motional ancillae approach can be further developed into a fully functional qubit register. Utilizing motional modes in ions for certain special applications is already well underway \cite{Zahringer2010,Toyoda2015,Chen2023,Whitlow2023,davoudi2021toward,navickas2024experimental}, with ongoing efforts realizing motional qubits capable of readout and performing general unitary operations \cite{hou2024coherent}. Similar to the all-to-all connectivity available with entangling gates, these ancillary states can couple with nearly any qubit in the system register. This capability holds the potential to create arbitrary probability distributions for general thermal states---using, e.g., circuits inspired by autoregressive models \cite{liu2021solving}. By leveraging otherwise unused degrees of freedom, our proof-of-concept can also be adapted to bosonic modes in other quantum systems, such as cavity quantum electrodynamics and superconducting circuit platforms.

The use of motional ancillae for the generation of thermal states provides a practical toolbox for studying quantum many-body systems at finite temperature. Applicable to fields like condensed matter physics, chemistry, and particle physics, our results pave the way for leveraging quantum computing to explore phase diagrams and thermodynamic properties in gauge theories and beyond.

\begin{acknowledgments}
%DoE wants this sentence to come first:
This material is based upon work supported by the U.S. Department of Energy (DoE), Office of Science, National Quantum Information Science Research Centers, Quantum Systems Accelerator (DE-FOA-0002253, NML). 
NML acknowledges support from the DoE Early Career Research Program (DE-SC0024504, NML). Additional support is acknowledged from the National Science Foundation, Quantum Leap Challenge Institute for Robust Quantum Simulation (OMA-2120757, NML) and Software-Tailored Architecture for Quantum Co-Design (STAQ) Award (PHY-2325080, NML). 
This research was also supported by the Natural Sciences and Engineering Research Council of Canada (NSERC, RL, CM), the Canada First Research Excellence Fund (CFREF Transformative Quantum Technologies, CM), Ontario Early Researcher Award (CM), and the Canadian Institute for Advanced Research (CIFAR, CM).
The authors thank Yunseong Nam for help with the quantum circuit design and helpful discussions, and Yingyue Zhu for experimental support.
\end{acknowledgments}

\section*{Author Contributions Statement}
A.T.T. transpiled and optimised quantum circuits to run on hardware, set up the experimental workflow, and analysed data.  A.M.G. developed the idea of motional ancillae for thermal state preparation.  K.W. integrated the VQE optimization into the experimental workflow.  A.T.T., M.T.D., X.L., and A.M.G. collected data.  A.M.G. and N.M.L. supervised the experimental work.  N.M.L. and C.A.M. coordinated the collaboration.  Y.Y.A, A.C., J.Z., R.L., and C.A.M. developed the theory results together. Y.Y.A. and A.C. developed the VQE protocol, performed numerical analysis, and adapted the group theoretical projection technique for non-Abelian lattice gauge theories. C.A.M. supervised the theoretical work. All authors contributed to the manuscript.\\~\\

\section*{Competing Interests Statement}
N.M.L. is the chief technology officer of TAMOS Inc.  A.T.T. and A.M.G. are inventors in a patent application related to the realization of motional ancillae. The remaining authors declare no competing interests.\\

% \newpage
% \appendix
\renewcommand{\thesection}{}
\renewcommand{\thesubsection}{\Alph{subsection}}
\section*{Methods} 
%\phantomsection
\label{methods}%can be merged into combined section later

%\section{Experimental methods}
\subsection{Trapped-ion platform setup}
\label{app:exp_platform}

The experiment described here significantly expands the capabilities of our fully programmable ion-trap quantum computer ~\cite{debnath_demonstration_2016}. The system is based on a chain of $^{171}$Yb$^+$ ions confined in a linear Paul trap.  Each ion hosts a pseudo-spin qubit encoded in the hyperfine splitting of the electronic ground state, with $\ket{0} = \ket{^2S_{1/2}, F=0, m_F=0}$ and $\ket{1} = \ket{^2S_{1/2}, F=1, m_F=0}$. The qubit splitting is approximately 12.643 GHz.  Laser beams resonant with the $^2S_{1/2}\rightarrow^2P_{1/2}$ transition are used to initialize the qubit into $\ket{0}$ through optical pumping and to perform projective measurements through state-specific fluorescence~\cite{olmschenk2007manipulation}. The state of each qubit in the chain is measured individually by focusing the scattered light for each ion onto a distinct photomultiplier tube (PMT). Single qubit measurement fidelities are greater than 99\%, limited by off resonant coupling, a fundamental limitation to fluorescence-based state detection. Detector cross talk further contributes to lower multi-qubit measurement fidelities ranging from 92 to 99\%, depending on the state. To mitigate these two errors, we perform an independent characterization of state measurement. Using a single ion to eliminate cross talk, we synthetically create representative measurement signals of each multi-qubit state and determine the probability that such a state would be measured correctly or incorrectly. This process allows us to eliminate the effect of measurement cross talk and off-resonant coupling from the qubit probability measurements which underpin the energy measurements made in this work.

Coherent manipulation of the qubit state is driven by off-resonant Raman transitions using two counter-propagating pulsed laser beams at 355 nm~\cite{hayes2010entanglement}. These operations include a universal gate set consisting of arbitrary single qubit rotations and all-to-all connected $R_{XX}^{i,j}(\theta) = \exp(-i\theta\, \hat{\sigma}^x_i \otimes\hat{\sigma}^x_j/2)$ entangling gates, utilising the M{\o}lmer-S{\o}rensen (MS) interaction~\cite{sorensen1999quantum}. Single- and two-qubit fidelities are greater than 99.9 and 98\%, respectively. Of the two beams needed to drive the Raman transition, one beam is split into an array of individual addressing beams, such that each unique beam has independent frequency, phase, and amplitude control and is focused on one ion. The second beam illuminates the chain as a whole for simplicity. The MS gates are implemented using pulse shaping techniques in~\cite{blumel2021efficient}.

The radial modes along the $x$-axis are chosen to mediate gates while a subset of the radial modes along the $y$-axis are used as ancillae (see Fig.~\ref{fig:protocol}), so there is no interference between them.

\subsection{Thermal state preparation using motional ancillae}
\label{app_thermal}

The first step to preparing each system qubit-ancilla pair in a suitable arbitrary superposition is to initialize to $\ket{\mathrm{spin,motion}}=\ket{0,0}$. After the motion is laser cooled to near the Doppler limit, all ions in the chain are subject to an optical pumping beam which places them in $\ket{\mathrm{spin}}=\ket{0}$ with a fidelity greater than 99.5\% in 5 microseconds. Subsequently, we perform resolved sideband cooling on all motional modes, with each mode requiring approximately 200 microseconds to reach the ground state~\cite{monroe1995resolved}. 

\begin{figure}[!h]
    \centering
    \includegraphics[width=1\linewidth]{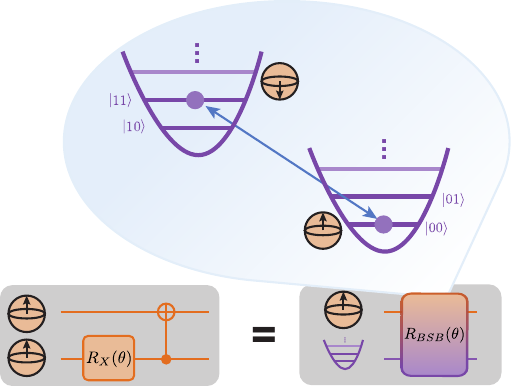}
    \caption{Two methods for thermal state preparation. Bottom: The protocol suitable for a qubit ancilla using a CNOT gate. Top: Alternative protocol utilising a motional ancilla. The system qubit and ancillary motional mode are both in their ground states. A blue sideband transition on the qubit resonance coherently transfers population from $\ket{\mathrm{qubit,motion}}=\ket{0,0}$ to $\ket{1,1}$, resulting in the final state $\cos(\theta/2)\ket{0,0}+\sin(\theta/2)\ket{1,1}$.} 
    \label{fig:BSBStatePrep}
\end{figure}

After all modes are prepared in the ground state, the ion chain's motion is in the Lamb-Dicke regime with respect to the Raman laser beams so the laser may be frequency tuned to drive a resonant blue sideband (BSB) transition on any motional mode~\cite{leibfried2003quantum}. Hence, for each qubit-mode pair, population can be coherently transferred between $\ket{0,0}$ and $\ket{1,1}$, such that the the final state is $\cos(\theta/2)\ket{0,0}+\sin(\theta/2)\ket{1,1}$ (see Fig.~\ref{fig:BSBStatePrep}). Here, $\theta=\Omega_{\mathrm{BSB}}\tau=\eta_{i,m}\Omega_{0}\tau$ with $\tau$ being the gate time, $\Omega_0$ being the Rabi frequency of the qubit transition, and $\eta_{i,m}$ being the Lamb-Dicke parameter for ion $i$ and mode $m$. Because these BSB pulses are used to create incoherent superpositions, a strict determination of their fidelity is not necessary. In fact, for the purpose of thermal state preparation, only the relative population in $\ket{0,0}$ and $\ket{1,1}$ is relevant. This ratio can be prepared with about 98\% accuracy.

Each motional ancilla may be assigned arbitrarily to almost any system qubit, as long as $\eta_{i,m}$ is not impractically small~\cite{kang2023efficient}. First, we describe the choice of qubit-mode pairs for the experiment presented in Fig.~\ref{fig:targetplot}. Although this experiment strictly requires only as many ions as there are system qubits, we choose the number of ions in the chain to not only support the required number of qubits, but also to optimise alignment of each ion to its addressing optics. For the experiments represented in Fig.~\ref{fig:targetplot}, we use a chain of seven ions, with four hosting system qubits. Of the seven radial modes along the $y$-axis, the second, fourth, and sixth modes are not used as ancillae, in order to limit deleterious off-resonant driving arising from imperfect mode resolution. Here, we index the ions from one to seven according to their position in the chain and the modes from one to seven, with mode one being the highest energy (center of mass) mode. 

On top of the mode resolution consideration, we choose the set of qubit-motional mode pairs with generally higher values of $\eta_{i,m}$ to minimise the time needed for state preparation. Specifically, we choose the pairs designated by $\eta_{i,m}=\{\eta_{2,1},\eta_{3,3},\eta_{4,5},\eta_{5,7}\}$. The motional modes used as ancillae for ions $\{2-5\}$ have radial frequencies $\omega=2\pi\times \{2.893,2.863,2.830,2.786\}$ MHz.

For the experiment shown in Fig.~\ref{fig:targetplot2}, we perform our experiment with nine ions in the trap. Ions $3-8$ host pseudo-spin system qubits. These are paired with modes designated by $\eta_{i,m}=\{\eta_{3,6},\eta_{4,5},\eta_{5,2},\eta_{6,9},\eta_{7,8},\eta_{8,7}\}$. The motional modes used as ancillae have radial frequencies $\omega=2\pi\times \{2.840,2.854,2.889,2.788,2.807,2.824\}$ MHz.

\subsection{Noise model for VQE}\label{app:noise_model}
To evaluate the performance of our VQE ansatz in the presence of noise, we used a device-aware noise model. The two primary noise sources in our trapped-ion quantum computer are: (i) random over- or under-rotations in the angles of the partial sideband gate coupling the ancillae with the system, and (ii) imperfections in the implementation of the MS gate on the system register.

Given a target rotation $R_X(\theta_i) = \exp(-i\theta\, \hat{\sigma}^x_i/2)$ on a ancilla qubit, we model the rotation error by applying $R_X(\theta_i')$ on the ancilla mode, where $\theta_i'$ is sampled from a normal distribution $\mathcal{N}(\theta_i,0.03\times\theta_i)$ for each cost function evaluation. The factor of 0.03 in the standard deviation is specific to our device's characteristics. Consequently, we substitute $\theta_i'$ for $\theta_i$ in our density matrix and entropy calculations.

The noisy MS gate in the system register is simulated using a two-qubit depolarizing noise channel applied after each MS gate. The channel strength is chosen to ensure an MS gate fidelity of 98\%, consistent with the experimental fidelity achieved in our system. We note that an MS gate fidelity greater than 95\% captures the phase transition shown in Figs.~\ref{fig:targetplot}c and \ref{fig:targetplot2}c well.

The statistical error introduced by the projective measurement is modeled by sampling the eigenvalues of our observables with $N_{meas}=2000$ shots for SU$(2)$ and $N_{meas}=3000$ shots for SU$(3)$, respectively.  The variational search was conducted using the PyBADS optimiser \cite{singh2024pybads,acerbi2017practical}, with a maximum of 230 function evaluations for SU(2) gauge theory and 350 for SU(3).

For each chemical potential $\mu$, we conducted 20 independent noisy simulations. In each run, we evaluated 10 instances of the thermal expectation value of the chiral condensate using our charge-singlet measurement protocol. The average of these 10 measurements provided a single data point, resulting in 20 averages per $\mu$ value. To quantify the accuracy of the measurements, we used a box plot and calculated the interquartile range (IQR), shown as the grey boxes in Fig.~\ref{fig:targetplot}c. The error bar represents the spread of data points, and in the presence of outliers denotes the interval (Q$_1-1.5$ IQR, Q$_3+1.5$ IQR), where Q$_1$ and Q$_3$ represents the first and third quartile of the data set.
This comprehensive analysis allowed us to evaluate the robustness of the VQE circuit and accuracy of our gauge-invariant measurement protocol under realistic noise conditions.

\subsection{Derivation of Eq.~(\ref{eq:projection_formula})} \label{app:projection_operator}
The projection operator enables charge-singlet measurements of observables in our protocol on states that are not necessarily charge-singlet. We use it to compute averages of operators within the singlet subspace, defined by the color-neutrality constraint. Our approach adapts and extends the ideas presented in \cite{elze1986quantum,leyaouanc1989}, tailoring them for use and implementation on a quantum computer. 

We start with the general formalism for an SU$(N_c)$ gauge group, where $N_c$ is the number of colors.
The Hilbert space of our system can be decomposed as $\mathcal{H} = \bigoplus_{\mathbf{\alpha}} \mathcal{H}_{\mathbf{\alpha}}$, where the direct sum runs over the irreducible representations $\alpha$ of the color group.    
Any gauge-invariant operator such as the Hamiltonian $\hat{H}$ or the Gibbs density operator $\hat{\rho}_{G}$ can be decomposed into a sum over irreducible representations and acts on the Hilbert space without mixing different representations. 
In particular, the color singlet subspace with $\alpha=0$ is of interest in this work. 

Following \cite{elze1986quantum}, we can express the trace over the whole Hilbert space of the product of a gauge-invariant observable $\hat{\Omega}$ with a general group element $\hat{U} \in $ SU$(N_{c})$  as 
\begin{equation}
    \mathrm{Tr}(\hat{\Omega}\hat{U}) = \sum_{\alpha} \frac{\mathrm{Tr}_{\alpha}(\hat{\Omega})\mathrm{Tr}_{\alpha}(\hat{U})}{d_{\alpha}},
\end{equation}
where the sum runs over irreducible representations of the color group, $\mathrm{Tr}_{\alpha}$ are traces restricted to states that transforms under the representation $\alpha$ and $d_{\alpha}$ is the dimension of the representation. 
In order to extract the trace over a particular representation, we make use of the orthogonality relation between the irreducible character functions $\chi_{\alpha}(\mathbf{\eta})=\mathrm{Tr}_{\alpha}\, \hat{U}(\eta)$ with respect to the Haar measure $d\mu(\eta)$ of the group $\int_{SU(N_c)} d\mu(\eta) \, \chi_{\alpha}^{\ast}(\mathbf{\eta})\chi_{\beta}(\mathbf{\eta})=\delta_{\alpha \beta}$, where $\eta$ are variables parametrizing group elements. 
Focusing on the charge-singlet subspace, for which the character function is given by $\chi_{0} =1$, we obtain the restricted trace 
\begin{equation}
    \mathrm{Tr}_{0}(\hat{\Omega}) =\mathrm{Tr}(\hat{\Omega} \hat{K}), \label{eq:method_restricted_trace}
\end{equation} 
where the general expression of the charge-singlet projector is
\begin{equation}
 \label{eq:methods_projector_general}
     \hat{K}=\int d\mu  \, \hat{U}.
 \end{equation}

Our charge-singlet measurement protocol is based on Eq.~(\ref{eq:method_restricted_trace}) and allows us to evaluate thermal averages restricted to the singlet  subspace from averages on the full Hilbert space. To see this, we first replace $\hat{\Omega}=\hat{\rho}_{G}$ with the Gibbs state $\hat{\rho}_{G}$ given in Eq.~(\ref{eq:gibbs_thermal}) and find that the thermal average of the operator $\hat{K}$ is equal to 
\begin{equation}
    \langle \hat{K} \rangle =\mathrm{Tr}(\hat{\rho}_{G} \hat{K})=\frac{Z_{0}}{Z} \;,
\end{equation}
where $Z_{0}=\mathrm{Tr}_{0} ( e^{-\beta \hat{H}}) $ is the gauge-single partition function and $Z$ the one over the full Hilbert space. By then choosing $\hat{\Omega}=\hat{O}\hat{\rho}_{G}$, where $\hat{O}$ is a physical observable,   we recover the charge-singlet measurement formula (\ref{eq:projection_formula}) in the main with $\langle \hat{O}\rangle_{0}=\mathrm{Tr}_{0} (\hat{O}e^{-\beta \hat{H}})/Z_{0}=\langle \hat{O}\hat{K}\rangle /\langle \hat{K}\rangle$.

Our formula is particularly well-suited for implementation on a quantum computer, as it requires only the measurement of two observables, $\hat{O}\hat{K}$ and $\hat{K}$, to recover the charge-singlet thermal average of the observable $\hat{O}$. To realise our charge-singlet measurement protocol, it is necessary to evaluate the projection operator $\hat{K}$.  The group integral in Eq.~(\ref{eq:methods_projector_general}) is explicitly computed for the SU$(2)$ and SU$(3)$ gauge group in Supplementary Information~\ref{app:projector_integration}.

In our protocol, we relegate the projection to the end of the VQE process rather than incorporating it into the VQE loop. Evaluating the projected energy during the VQE is resource-intensive. Additionally, it would require knowledge of the entropy within the singlet subspace to compute the cost function.

\subsection{Details of SU$(2)$ for experimental realisation}
\subsubsection{SU$(2)$ gauge group basic building block $N=2$}\label{app:su2_unit_cell}
\begin{figure*}[!th]
    \centering
    \includegraphics[width=1.0\linewidth]{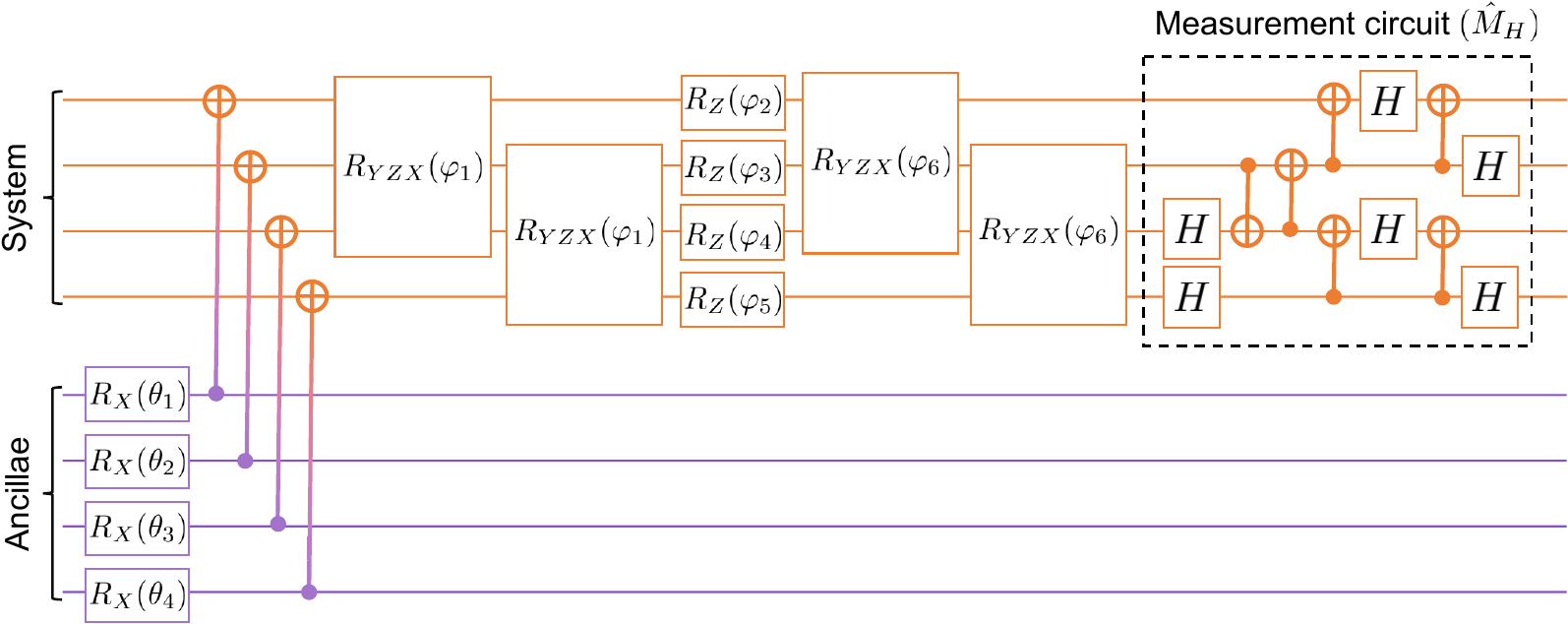}
    \caption{Circuit for the variational preparation of SU(2) LGT thermal states. The circuit includes parameterized $R_{X}(\theta_i)$ rotations applied to the ancillae, followed by CNOT gates coupling the motional ancillae with the system qubits. This first group of operations forms $\hat{U}_{A}(\theta)$. Then, a layer of $R_{Z}$ rotations sandwiched between blocks of three-body $R_{YZX}$ gates is applied. The gates acting on the system qubits form the unitary operation $\hat{U}_{S}(\varphi)$.  The circuit has 10 variational parameters. Additionally, a measurement circuit is required for measuring the non-diagonal contribution $\hat{H}_{2}$ in the Hamiltonian. } 
    \label{fig:SU2_circuit}
\end{figure*}

The general expression of the SU$(2)$ Hamiltonian for $N$ sites can be found in  Supplementary Information~\ref{sec:su2_general}.
Our experimental demonstration focuses on the unit cell with $N=2$ lattice sites consisting of two antimatter and matter fermions with (anti-)red and (anti-)green colors, which is mapped to a system of 4 qubits. The Hamiltonian $\hat{H}=\hat{H}_1+\hat{H}_2$ decomposes into two non-commuting families 
\begin{align}
    \notag \hat{H}_{1}=&\left(2m+\frac{3}{16x}\right)+ \left(\frac{m}{2}-\frac{\mu}{4}\right)(\hat{\sigma}_{3}^{z}+\hat{\sigma}_{4}^{z}) \\
    &-\left(\frac{m}{2}+\frac{\mu}{4}\right)(\hat{\sigma}_{1}^{z}+\hat{\sigma}_{2}^{z})-\frac{3}{16x}\hat{\sigma}_{1}^{z}\hat{\sigma}_{2}^{z}\,,
\end{align}
and 
\begin{equation}
    \hat{H}_{2}=-\frac{1}{4}(\hat{\sigma}_{1}^{x}\hat{\sigma}_{2}^{z}\hat{\sigma}_{3}^{x}+\hat{\sigma}_{1}^{y}\hat{\sigma}_{2}^{z}\hat{\sigma}_{3}^{y}+\hat{\sigma}_{2}^{x}\hat{\sigma}_{3}^{z}\hat{\sigma}_{4}^{x}+\hat{\sigma}_{2}^{y}\hat{\sigma}_{3}^{z}\hat{\sigma}_{4}^{y})\,,
\end{equation}
where $m$, $\mu$ and $x=1/g^2$ are the mass, chemical potential and inverse coupling constant respectively. $\hat{\sigma}_{n}^{i}$ with $i=x,y,z$ denotes the usual single qubit Pauli matrices at site $n$. $\hat{H}_1$ consists of exclusively diagonal Pauli strings and $\hat{H}_2$ contains only  non-diagonal Pauli strings.  Pauli operators within the same family commute with each other and can therefore be measured simultaneously.   However, since $\hat{H}_{2}$ is non-diagonal, a measurement circuit is required in practice to rotate it to the diagonal basis before performing measurements in the $\hat{\sigma}^{z}-$basis (see Fig.~\ref{fig:SU2_circuit}). For our target plot shown in Fig.~\ref{fig:targetplot}, we fix $m=0.5$, $x=1$, while the chemical potential $\mu$ varies from $0$ to $4$. The coefficients of the Pauli strings for $\hat{H}_{1}$ thus vary with the chemical potential $\mu$.

\subsubsection{SU$(2)$ VQE circuit}\label{app:su2_vqe_circ}
For the basic building block studied here, we need 4 system qubits and 4 motional ancilla modes.  
The circuit employed in the VQE protocol consists of two main parts.  First, a  parametrized unitary $\hat{U}_{A}(\boldsymbol{\theta})$ is applied to couple the ancillae with the system qubits
\begin{equation}
    \hat{U}_{A}(\boldsymbol{\theta})=\bigotimes_{i=1}^{4} R_{X}(\theta_{i})\,\mathrm{CNOT}_{A_{i},S_{i}}\,,
\end{equation}
where $R_{X}(\theta_{i})=\exp(-i\theta_{i} \, \hat{\sigma}_{i}^{x}/2)$ denotes the rotation around the $x$-axis by an angle $\theta_{i}$ on the ancilla mode $i$, and $\mathrm{CNOT}_{A_{i},S_{i}}$ entangle each mode $i$ in the ancilla register $\mathcal{A}$ with the qubit $i$ in the system register $\mathcal{S}$ (see Fig.~\ref{fig:SU2_circuit}). From here on, we will use the notation $R_P(\theta) = \exp(-i\theta \, \hat{P}/2) $ for a rotation gate, where $\hat{P}$ is a Pauli string.
By tracing out the ancilla modes, we obtain the system's density matrix
\begin{align}
    \hat{\rho}_S = {\rm Tr}_A(\hat{\rho}_{AS}) = \bigotimes_{i=1}^{4} \begin{pmatrix}
        \cos^2(\theta_i/2) & 0 \\
        0 & \sin^2(\theta_i/2)
    \end{pmatrix}.
\end{align}
Expanding this in the computational basis, the density matrix of the system is
\begin{align}
    \hat{\rho}_S = \sum_{j} \tilde{p}_{j} \ket{j}\bra{j}, \label{appendix:density_matrix}
\end{align}
where $j$ denotes the computational basis vectors $\ket{j} = \ket{j_1 j_2 j_3 j_4}$ with each $j_i \in \{0,1\}$. The entropy of the system is then analytically obtained from the probabilities $\tilde{p}_{j}$ of the bit string  $j$ using  Eq.~(\ref{eq:entropy_analytical}).

In the second part of the circuit, the state in Eq.~(\ref{appendix:density_matrix}) is evolved by the unitary $\hat{U}_S(\boldsymbol{\varphi})$ acting only on the system qubits to get the desired thermal state. The unitary gates in $\hat{U}_{S}$ are inspired by the Pauli strings appearing in the decomposition of the Hamiltonian. Specifically, we use the three-body gate 
\begin{equation}
    R_{YZX}(\varphi_{i})\equiv \exp\left(-i\varphi_{i} (\hat{\sigma}^{y}\otimes \hat{\sigma}^{z}\otimes \hat{\sigma}^{x})/2 \right).
\end{equation}
Here, we specifically choose $R_{YZX}$ gates instead of $R_{XZX}$, as their commutation with Pauli strings in $\hat{H}_{1}$ accurately reproduces the terms of the Hamiltonian in $ \hat{H}_{2}$.
This 3-body gate can be decomposed into native entangling MS gates (see Supplementary Information~\ref{app:circ_decomp}). In total, we need three entangling $R_{XX}$ gates to implement the three-body gates.  In our circuit design, we employ a shifted block structure, where we first apply two consecutive $R_{YZX}$ gates sharing the same variational parameters on consecutive three qubits, then  apply a layer of single qubit $R_{Z}$ with independent variational parameters and finally apply two additional parameter-sharing three-body gates. Using gate identities, the system circuit can be reduced to have only 8 MS gates compared to the initial naive counting of 18 MS gates. The reduced circuit in terms of native gates is shown in Fig.~\ref{fig:reduced_su2_circ} in Supplementary Information~\ref{app:circ_decomp}.

The circuit design outlined above is scalable and can be readily extended to larger lattice sizes by increasing the number of qubits in the ancilla and system registers. Because the many-body nature of the interactions in the Hamiltonian remains fixed across different lattice sizes, the type of the gates in the circuit also remains unchanged. The number of two-qubit gates scales polynomially with system size, as the number of Pauli strings in the Hamiltonian grows polynomially with $N$. The ancilla circuit is similarly straightforward to generalise for larger systems. However, at general values of temperature and chemical potential, preparing the thermal state may necessitate entangling operations among the motional ancilla modes. This, in turn, would require measurements on the motional ancillae to determine the entropy. Multiple techniques for measuring trapped-ion motional states have already been demonstrated in small systems, with their application to larger systems being limited by low motional coherence times~\cite{meekhof1996generation,um2016phonon,mallweger2023single,hou2024coherent}. Coherence time improvements driven by growing interest in quantum technology based on qumodes will render motional mode measurements feasible on larger devices~\cite{Chen2023}.

After the circuit execution, a measurement in the computational basis allows us to determine and measure the diagonal contribution of the Hamiltonian $\hat{H}_{1}$. 
Since the Hamiltonian decomposition also contains non-diagonal Pauli strings given by $\hat{H}_{2}$, we need to integrate an additional circuit $\hat{M}_{H}$ to the unitary $\hat{U}_{S}$ (indicated as measurement circuit in Fig.~\ref{fig:SU2_circuit}) in order to measure $\hat{H}_{2}$. To find the measurement circuit $\hat{M}_{H}$, we used the stabilizer approach to transform the stabilizer matrix associated with the commuting family of Pauli strings in $\hat{H}_{2}$ into its representation in the computational basis \cite{gokhale2019minimizing}. 
The circuit $\hat{M}_{H}$ diagonalizes the Hamiltonian $\hat{H}_{2}$.

\subsubsection{Charge-singlet measurement of $\hat{\chi}$ for SU$(2)$}\label{app:su2_chiral_cond}
The observable of interest in our study is the chiral condensate 
\begin{equation}
    \hat{\chi} = \sum_{n = 1}^{N} \frac{(-1)^{n}}{2}\left( \hat{\sigma}_{2n-1}^{z}  + \hat{\sigma}_{2n}^{z}\right)
\end{equation}
and serves as an order parameter to probe the phase diagram at finite temperature and chemical potential.
In order to evaluate its thermal average 
in the charge-singlet subspace, we use Eq.~(\ref{eq:projection_formula}) with $\hat{O}=\hat{\chi}$
 \begin{equation}
     \langle \hat{\chi}\rangle_{0}= \frac{\langle \hat{\chi}\hat{K}\rangle}{\langle \hat{K}\rangle},
 \end{equation}
 where $\langle \hat{\chi}\rangle_{0}=\mathrm{Tr}_{0}\{e^{-\beta \hat{H}} \hat{\chi} \}/Z_{0} $ and $Z_{0}=\mathrm{Tr}_{0}(e^{-\beta \hat{H}})$ is the singlet partition function.
The thermal averages on the right hand side are expressed in the full Hilbert space or the unconstrained space as $\langle \hat{O} \rangle = \mathrm{Tr} (\hat{\rho} \hat{O})$.

The group integral defining our projector in Eq.~(\ref{eq:methods_projector_general}) can be evaluated exactly for SU$(2)$. 
The general expression of the operator $\hat{K}$ in terms of the diagonal charge $\hat{Q}_{tot}^{z}$ can be found in Eq.~(\ref{eq:K_SU2_general}).
Since  $\hat{Q}_{tot}^{z}$ is a diagonal operator, the projection operator $\hat{K}$ is also diagonal in the computational basis. In particular, for $N=2$, the Pauli decomposition of the operator $\hat{K}$ reads
\begin{align}
    \notag \hat{K}=&\frac{3}{16}(\hat{\sigma}_{3}^{z}\hat{\sigma}_{4}^{z}+\hat{\sigma}_{2}^{z}\hat{\sigma}_{3}^{z}+\hat{\sigma}_{1}^{z}\hat{\sigma}_{4}^{z}+\hat{\sigma}_{1}^{z}\hat{\sigma}_{2}^{z}-\hat{\sigma}_{2}^{z}\hat{\sigma}_{4}^{z}-\hat{\sigma}_{1}^{z}\hat{\sigma}_{3}^{z}) \\
    & +\frac{5}{16}(1+\hat{\sigma}_{1}^{z}\hat{\sigma}_{2}^{z}\hat{\sigma}_{3}^{z}\hat{\sigma}_{4}^{z}),
\end{align}
and 
\begin{align}
    \notag \hat{\chi}\hat{K}=-\frac{1}{4}(&\hat{\sigma}_{1}^{z}+\hat{\sigma}_{2}^{z}+\hat{\sigma}_{2}^{z}\hat{\sigma}_{3}^{z}\hat{\sigma}_{4}^{z}+\hat{\sigma}_{1}^{z}\hat{\sigma}_{3}^{z}\hat{\sigma}_{4}^{z}\\
    & -\hat{\sigma}_{3}^{z}-\hat{\sigma}_{4}^{z}-\hat{\sigma}_{1}^{z}\hat{\sigma}_{2}^{z}\hat{\sigma}_{3}^{z}-\hat{\sigma}_{1}^{z}\hat{\sigma}_{2}^{z}\hat{\sigma}_{4}^{z}
    ).
\end{align}

In practice, after the VQE optimisation concludes and the optimal parameters $(\boldsymbol{\theta}^{\star},\boldsymbol{\varphi}^{\star})$ are found, they are used to evaluate the expectation values of the observables  $\hat{\chi}\hat{K}$ and  $\hat{K}$ on the quantum hardware. Since both observables are diagonal in the computational basis, no additional quantum resources are required for their measurement. 

\subsection{Details of SU$(3)$ for experimental implementation}
\begin{figure*}[t] 
    \centering
    \includegraphics[width=0.96\textwidth]{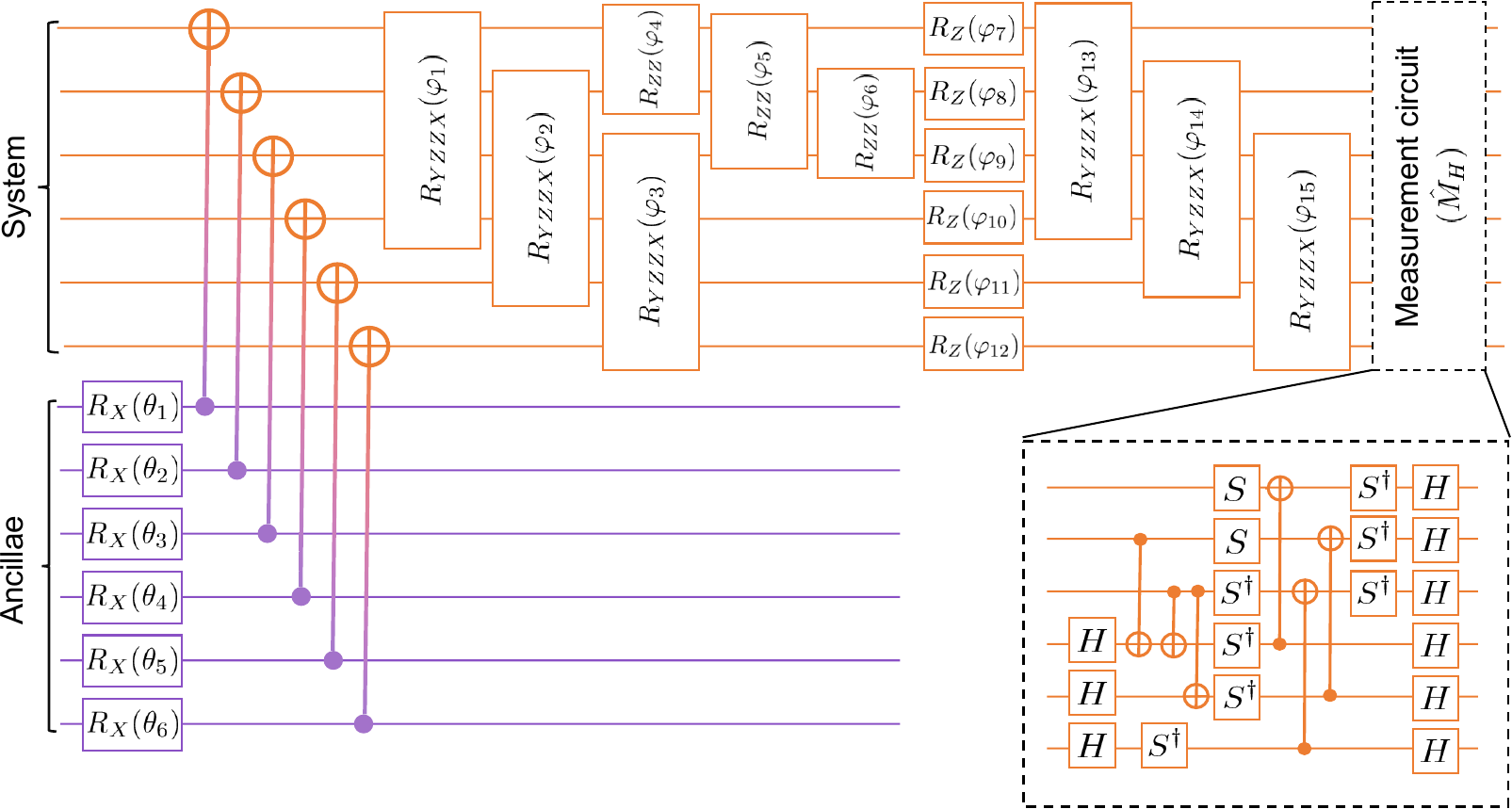} 
    \caption{Circuit for preparing the SU(3) thermal state of a unit cell with $N=2$. The circuit includes parameterized $R_{X}(\theta)$ rotations applied to the ancilla qubits, followed by a series of CNOT gates that entangle the ancilla qubits with the system qubits. Post-entanglement, a series of three four-body gates 
    $R_{YZZX}(\varphi)$ with independent variational parameters is applied, followed by three two-body $R_{ZZ}$ gates and a layer of parametrized $R_{Z}$ rotations. The unitary $\hat{U}_{S}(\boldsymbol{\varphi})$ concludes with another series of three four-body gates. A measurement circuit $\hat{M}_{H}$, shown in the inset, is required for measuring the non-diagonal contribution $\hat{H}_{2}$ in the Hamiltonian. In total, $21$ variational parameters are needed for the simulation. }
\label{fig:methods_su3_circuit}
\end{figure*}

\subsubsection{SU$(3)$ basic building block for $N=2$}\label{app:su3_unit_cell}
In this work, we perform a finite temperature VQE for the basic building block ($N=2$) of SU$(3)$. The general expression of the Hamiltonian for $N>2$ can be found in the Supplementary Information~\ref{sec:su3_general}. For $N=2$, the Hamiltonian describes a system of $6$ qubits. 
Written in terms of the Pauli operators, the Hamiltonian $\hat{H}=\hat{H}_1+\hat{H}_2$ decomposes into two non-commuting families given by 
\begin{align}
    \notag \hat{H}_{1}=&\left( \frac{m}{2}-\frac{\mu}{6}\right)(\hat{\sigma}_{4}^{z}+\hat{\sigma}_{5}^{z}+\hat{\sigma}_{6}^{z}) -\left( \frac{m}{2}+\frac{\mu}{6}\right)(\hat{\sigma}_{1}^{z}+\hat{\sigma}_{2}^{z}\\
    & +\hat{\sigma}_{3}^{z})-\frac{1}{6x}(\hat{\sigma}_{1}^{z}\hat{\sigma}_{2}^{z}+\hat{\sigma}_{1}^{z}\hat{\sigma}_{3}^{z}+\hat{\sigma}_{2}^{z}\hat{\sigma}_{3}^{z})+\left(3m+\frac{1}{2x}\right), \label{eq:methods_SU3_H1}
\end{align}
\begin{align}
   \notag  \hat{H}_{2}=\frac{1}{4}(&\hat{\sigma}_{2}^{x}\hat{\sigma}_{3}^{z}\hat{\sigma}_{4}^{z}\hat{\sigma}_{5}^{x}+\hat{\sigma}_{2}^{y}\hat{\sigma}_{3}^{z}\hat{\sigma}_{4}^{z}\hat{\sigma}_{5}^{y}-\hat{\sigma}_{1}^{x}\hat{\sigma}_{2}^{z}\hat{\sigma}_{3}^{z}\hat{\sigma}_{4}^{x}-\hat{\sigma}_{1}^{y}\hat{\sigma}_{2}^{z}\hat{\sigma}_{3}^{z}\hat{\sigma}_{4}^{y} \\
    & -\hat{\sigma}_{3}^{x}\hat{\sigma}_{4}^{z}\hat{\sigma}_{5}^{z}\hat{\sigma}_{6}^{x}-\hat{\sigma}_{3}^{y}\hat{\sigma}_{4}^{z}\hat{\sigma}_{5}^{z}\hat{\sigma}_{6}^{y}).
\end{align}

\subsubsection{SU$(3)$ VQE circuit}\label{app:su3_vqe_circ}
The SU(3) circuit used has the same structure as the one devised for the SU(2) gauge group. First, a layer of single qubit $R_{X}(\theta_{i})$ rotations with  $i=1,2,\dots,6$ is applied to the ancilla modes which are then independently coupled to the system qubit by a series of CNOT gates (see Fig.~\ref{fig:methods_su3_circuit}). The system qubits are then acted on with 4-body $R_{YZZX}$  gates inspired by the Pauli strings decomposition of the Hamiltonian 
\begin{equation}
    R_{YZZX}(\varphi_{i})\equiv \exp\left(-i\varphi_{i} (\hat{\sigma}^{y}\otimes \hat{\sigma}^{z}\otimes \hat{\sigma}^{z}\otimes \hat{\sigma}^{x})/2 \right).
\end{equation}
Again, here we use $R_{YZZX}$ gates instead of $R_{YZZY}$ to recover the different terms of the Hamiltonian through the commutation algebra. 
The layer of four-body gates is followed by a series of two-body $R_{ZZ}$ gates. Next, a layer of single-qubit parametrized rotation gates $R_Z(\theta_i)$ is applied. The circuit concludes with another series of three four-body gates. Each four-body gate can be decomposed into five two-qubit gates. To measure the expectation value of the non-diagonal family of Pauli strings appearing in $\hat{H}_2$, a measurement circuit $\hat{M}_H$ is added to the system register (see Fig.~\ref{fig:methods_su3_circuit}).

The naive transpilation of this circuit into native gates results in 39 entangling MS gates in the system circuit (including the measurement circuit). 
However, this number can be reduced to 9 by using gate identities and circuit simplification techniques (see Fig.~\ref{fig:reduced_su3_circ} in Supplementary Information~\ref{app:circ_decomp}). The circuit implemented on the quantum hardware and used in the numerical simulation is the optimised version obtained after reduction.

\subsubsection{ SU$(3)$ chiral condensate charge-singlet measurement}\label{app:su3_chiral_cond}
A general group element $\hat{U}\in \mathrm{SU}(3)$ can be parametr-ized using eight variables $\lbrace \eta_{a}\rbrace_{a=1,\dots,8}$ and the eight non-Abelian charges $\hat{Q}_{tot}^{a}$ (see Supplementary Information~\ref{sec:su3_general})
\begin{equation}
    \hat{U}=\exp \left( i \sum_{a=1}^{8}\eta_{a}\hat{Q}_{tot}^{a} \right).
\end{equation}

The SU$(3)$ group possesses two diagonal charge generators $\hat{Q}_{tot}^{3}$ and $\hat{Q}_{tot}^{8}$.  We can thus calculate the projection operator $\hat{K}$ by computing a double integral over the diagonal charges (see Supplementary Information~\ref{app:projector_integration} for more details). In particular, for the unit cell with $N=2$, the Pauli decomposition of the projector $\hat{K}$ reads 
\begin{align}
\hat{K} & =
\frac{5}{96} \sum_{i<j} \hat{\sigma}_i^z \hat{\sigma}_j^z + \frac{1}{96} \sum_{i<j<k<l} \hat{\sigma}_i^z \hat{\sigma}_j^z \hat{\sigma}_k^z \hat{\sigma}_l^z +\frac{3}{32} \, \hat{\mathbb{1} } \notag \\
&+\frac{1}{6}(\hat{\sigma}_1^z \hat{\sigma}_2^z\hat{\sigma}_4^z\hat{\sigma}_5^z +\hat{\sigma}_1^z \hat{\sigma}_3^z\hat{\sigma}_4^z\hat{\sigma}_6^z +\hat{\sigma}_2^z \hat{\sigma}_3^z\hat{\sigma}_5^z\hat{\sigma}_6^z -\hat{\sigma}_1^z\hat{\sigma}_4^z \notag \\
& \qquad -\hat{\sigma}_2^z\hat{\sigma}_5^z - \hat{\sigma}_3^z\hat{\sigma}_6^z ) - \frac{1}{32} \, \hat{\sigma}_1^z \hat{\sigma}_2^z \hat{\sigma}_3^z \hat{\sigma}_4^z \hat{\sigma}_5^z \hat{\sigma}_6^z, \label{eq:methods_K_su3}
\end{align}
where $i,j,k,l=1,2,\dots,6$.
The general formula for the SU(3) chiral condensate is given by

\begin{equation}
    \hat{\chi} = \sum_{n = 1}^{N} \frac{(-1)^{n}}{2}\left( \hat{\sigma}_{3n-2}^{z}  + \hat{\sigma}_{3n-1}^{z}+\hat{\sigma}_{3n}^{z}\right).
\end{equation}
For the unit cell, $N=2$, the chiral condensate decomposes as $\hat{\chi} = (-\hat{\sigma}_{1}^{z}-\hat{\sigma}_{2}^{z}-\hat{\sigma}_{3}^{z}+\hat{\sigma}_{4}^{z}+\hat{\sigma}_{5}^{z}+\hat{\sigma}_{6}^{z})/2$. To obtain the decomposition of $\hat{\chi} \hat{K}$  necessary for the evaluation of $\langle\hat{\chi}\rangle_0$, we can multiply the two Pauli decompositions above. Similar to the SU$(2)$ case, the operators $\hat{\chi}\hat{K}$, and $\hat{K}$ are also diagonal, and can be measured on the quantum device without requiring an additional measurement circuit. 

\putbib[biblio]
\end{bibunit}
%\bibliography{biblio}{}

\clearpage
\begin{bibunit}
\widetext
\begin{center}
\textbf{\large Supplementary Information}
\end{center}
%%%%%%%%%% Merge with supplemental materials %%%%%%%%%%
%%%%%%%%%% Prefix a "S" to all equations, figures, tables and reset the counter %%%%%%%%%%
\setcounter{section}{0}
\setcounter{equation}{0}
\setcounter{figure}{0}
\setcounter{table}{0}
\makeatletter
\renewcommand{\theequation}{S\arabic{equation}}
\renewcommand{\thefigure}{S\arabic{figure}}
\renewcommand{\thesection}{S.\Roman{section}}
\renewcommand{\bibnumfmt}[1]{[S#1]}
\renewcommand{\citenumfont}[1]{S#1}

\section{Experimental details}
\subsection{Classical optimisation}
The free energy was minimised using Bayesian adaptive direct search (BADS) \cite{acerbi2017practical} via the PyBADS library in Python \cite{singh2024pybads}.  PyBADS alternates between a fast Bayesian optimisation and a slower, mesh-based exploration of the parameter landscape.  BADS is particularly suited for optimisation of noisy black-box functions with up to approximately twenty parameters.

For SU(2), we minimised the free energy for $\mu \in \{0.01,0.75,1.5,2.5,3.5\}$ and calculated the chiral condensate from the resulting states.  We started with the highest value of $\mu=3.5$ and performed VQE with a random initializiation to obtain optimal parameters.  To minimise the free energy of the next-highest value of $\mu=2.5$, we performed VQE initialized with the optimal parameters from $\mu=3.5$.  We repeated this process until the free energy was minimised for all 5 $\mu$ values.

For SU(3), we minimised the free energy for $\mu = 2.0$ by performing VQE initialized with optimal parameters obtained from a classically simulated VQE run for $\mu=2.25$.

\subsection{optimisation tuning}
PyBADS can be configured to start its optimisation run at a specific mesh size.  It will end the run when the mesh size is below a certain size, or when it reaches some number of function evaluations, both of which can be set by the user.  VQE runs on a simulated noisy system were used to determine the initial mesh size and maximum number of function evaluations.  The terminating mesh size was chosen to correspond with the experimental precision of the gate parameters.

For SU(2), we began with a mesh size of 1, and terminated after reaching a mesh size of 0.01 or 230 function evaluations.  For SU(3), we began with a mesh size of 0.25 and terminated after reaching a mesh size of 0.01 or 350 function evaluations.

\subsection{Gate calibration procedure}

Two experimental parameters affecting the accuracy of the quantum operations in this protocol drift significantly over the course of the VQE and therefore must be routinely calibrated. These are the frequency of the motional modes and the intensity of the Raman laser beams at the ions, which drops as the laser beams become slightly misaligned. While both these factors are checked for large drifts and corrected every few hours, we use active feedback on more dynamic parameters to compensate for the effect of small drifts.

Approximately every twenty minutes, we pause the VQE optimisation and perform calibration gates which should have a distinct level of rotation in the Bloch sphere. For the spin-only gates needed for $\hat{U}_S(\boldsymbol{\phi})$ and $\hat{M}_{H}$, the power in the gate lasers is adjusted such that the calibration gates achieve the correct level of rotation. For the spin-motion gates needed to prepare the thermal state, we adjust the duration of the sideband pulse.

\section{Evaluation of $\hat{K}$ for SU(2) and SU(3)}
\label{app:projector_integration}
In this section, we derive the projection operator for the singlet subspace. We begin with the general formalism for an SU$(N_c)$ gauge group, where $N_c$ is the number of color components. Then, we specialize to the $N_{c}=2$ and $N_c=3$ case to illustrate the construction of $\hat{K}$ explicitly.
The general expression of the projection operator for the singlet subspace is given by 
\begin{equation}
    \hat{K}=\int_{SU(N_{c})} d\mu  \, \hat{U}. \label{eq:K-nondiag}
\end{equation}
The projection operator in Eq.~(\ref{eq:K-nondiag}) is not diagonal in the computational basis because the group elements $U$ are generated by $N^2-1$ generators, not all of which are diagonal. A density matrix in the full Hilbert space can be projected onto the singlet subspace by the operation $\hat{K}\hat{\rho} \hat{K}$ properly normalized by the factor ${\rm Tr}(\hat{\rho}\hat{K})$. However, this operation is resource-extensive on the quantum computer due to the possible presence of exponentially many non-diagonal Pauli strings in the decomposition of $\hat{K}$. However, we are only interested in evaluating observable expectation values on the singlet subspace, not preparing the charge-singlet density matrix. In this context, the projection operator always appear within a trace operation.
Any group element $\hat
{U}$ in Eq.~(\ref{eq:K-nondiag}) can be diagonalized using a $SU(N_{c})$ rotation. Since we are only evaluating $\text{Tr}(\hat{\Omega}\hat{K})$, where $\hat{\Omega}$ is a gauge-invariant operator, we can 
choose and calculate $\hat{K}$ in the diagonal basis for simplicity \cite{mclerran1985thermodynamics}.  The justification of using only a parametrization of the Cartan subgroup for the projection operator is further demonstrated in \cite{elze1986quantum}. We  can thus reduce the integration over the whole SU$(N_{c})$ group to its maximal torus i.e., the largest subset of the group where elements commute \cite{greiner2012quantum}
\begin{equation}
    \hat{K} = \int_{SU(N_c)} d\mu(\eta_a) \; \exp\left(-i\sum_{a \in \mathcal{C}}\eta_a \hat{Q}_{tot}^a\right) \;.\label{eq:K_def_cartan}
\end{equation}
The sum in the exponential runs over the Cartan subalgebra $\mathcal{C}$ of the SU$(N_c)$ group defined by the conserved diagonal charges $\hat{Q}_{tot}^a$ and $d\mu(\eta_a)$ is the Haar measure of the group.

Let us now evaluate the explicit expression of the projector for SU$(2)$.
A general group element $\hat{U} \in $ SU$(2)$ can be parametrized with three angular variables
\begin{equation}
    \hat{U}=\exp \left( i \eta_{x}\hat{Q}_{tot}^{x}+i\eta_{y}\hat{Q}_{tot}^{y}+ i\eta_{z}\hat{Q}_{tot}^{z} \right),
\end{equation}
where the non-Abelian charges are given by Eq.~(\ref{appendix:su2_charges_x})-(\ref{appendix:su2_charges_z}). 
SU$(2)$ possesses only one diagonal charge $\hat
{Q}_{tot}^{z}$, and thus only one angular variable $\eta\in [0,4\pi]$ is needed to parametrize elements of the Cartan subspace.  
The Haar measure on SU(2) is expressed in terms of $\eta$ as
\begin{equation}
    d\mu= \frac{1}{2\pi}\sin^{2} \left( \frac{\eta}{2}\right)d\eta,
\end{equation}
where $\sin^{2} (\eta/2)$ arises from the Weyl measure (it is the Jacobian of the transformation) and the factor $1/2\pi$ ensures proper normalisation  \cite{blau11994lectures}. Using Eq.~(\ref{eq:K_def_cartan}), the projection operator $\hat{K}$ simplifies to
\begin{equation}
    \hat{K}=\frac{1}{2\pi}\int_{0}^{4\pi}\sin^{2} \left( \frac{\eta}{2}\right) e^{i\eta \hat{Q}_{tot}^{z}} d\eta = \frac{e^{2i\pi \hat{Q}_{tot}^{z}}\sin (2\pi \hat{Q}_{tot}^{z})}{2\pi \hat{Q}_{tot}^{z}(1-( \hat{Q}_{tot}^{z})^2) }. \label{eq:K_SU2_general}
\end{equation}
Since  $\hat{Q}_{tot}^{z}$ is a diagonal operator, the projection operator $\hat{K}$ is also diagonal in the computational basis. Note that the operator $\hat{K}$ is not singular as the numerator and denominator approach zero (which can happen for the eigenvalues $0$ and $\pm 1$ of $\hat{Q}_{tot}^{z}$) and remains well defined in this limit.

The SU(3) group possesses two diagonal charge generators $\hat{Q}_{tot}^{3}$ and $\hat{Q}_{tot}^{8}$ given in Eq.~(\ref{qubit_nonabeliancharge3}) and (\ref{qubit_nonabeliancharge8}). The maximal torus of SU(3) is thus isomorphic to $U(1) \times U(1)$ and any diagonal SU(3) matrix can be parametrized using two independent angles $(\eta, \psi) \in [-\pi,\pi]$ . The Haar measure on SU(3) can be written as \cite{greiner2012quantum}
\begin{equation}
    d\mu = \frac{8}{3\pi^{2}} \left(\sin(\eta)  \sin \left( (3\psi +\eta)/2\right) \sin \left((3\psi -\eta)/2\right)\right)^2 d \eta d\psi 
\end{equation}
and the projector is numerically obtained by computing the double integral
\begin{equation}
    \hat{K}=\int_{-\pi}^{\pi}\int_{-\pi}^{\pi} d \mu(\eta,\psi)\, e^{2i\eta \hat{Q}_{tot}^{3}}\, e^{2i\sqrt{3}\psi \hat{Q}_{tot}^{8}}.
\end{equation}

\section{SU(2) Hamiltonian in (1+1)D}
\label{sec:su2_general}

The Hamiltonian studied in this work is the discretised version of the one dimensional continuum Yang-Mills Hamiltonian defined on a spatial lattice whose points are separated by a length $a$ \cite{jackiw1980introduction}. We specifically adopt the staggered formulation of Kogut
and Susskind, where fermions and antifermions occupy
separate lattice sites, and are arranged in an alternating
pattern along the lattice (see Fig.~\ref{fig:protocol}a). The explicit form of the Hamiltonian for $N$ physical sites is \cite{kogut_hamiltonian_1975}
\begin{equation}
		\hat{H_l} = \frac{1}{2a} \sum_{n=1}^{N-1} \left( \hat{\phi}_{n}^\dagger \hat{U}_{n} \hat{\phi}_{n+1} + \operatorname{H.C.}\right)\\  +m \sum_{n=1}^{N} (-1)^{n} \hat{\phi}_{n}^\dagger \hat{\phi}_{n}  + \frac{a g^{2}}{2} \sum_{n=1}^{N-1} \hat{\boldsymbol{L}}_{n}^{2}-\mu \sum_{n=1}^{N}\hat{\phi}_{n}^\dagger \hat{\phi}_{n}
        \label{SIeq:KSham}
\end{equation}
where the matter field at each lattice site $n$ is described by a two-component fermionic field $\hat{\phi}_{n}=(\hat{\phi}_{n}^{1},\hat{\phi}_{n}^{2})^{T},$
and the upper index labels the two possible colors. Here we will focus on the SU(2) gauge group as an example but the generalisation to SU(3) is straightforward (see \cite{atas2023} for further details). The fermion mass is
denoted by $m$, $g$ quantifies the matter-field coupling constant and $\mu$ is the fermion chemical potential. 

The first term in the Hamiltonian corresponds to the kinetic energy and contains the parallel transporter (or connection) $\hat{U}_n=\exp(i\hat{\Omega}_{n}^{a}T^{a})$ that acts on the link between sites $n$ and $n+1$ and mediates the interaction between the internal color degree of freedom of the fermions on neighbouring sites. Its presence ensures the invariance of the Hamiltonian under local gauge transformations.  The $T^{a}=\sigma^{a}/2$ are the three generators of the $\operatorname{SU}(2)$ Lie algebra, and ${\sigma}^{a}$ the $a$-th Pauli matrix ($a=x,y,z$).
The angular variables $\hat{\Omega}_{n}^{a}$ are related to the continuum spatial component of the gauge field on the link  $n$ at position $z$ as $\hat{\Omega}_{n}^{a}/(ag) \rightarrow \hat{A}_{1}^{a}(z)$ when the lattice spacing goes to zero. Note that we adopt the temporal Weyl gauge where $\hat{A}_{0}^{a}(z)=0$ (see \cite{atas2021} for more details on the passage from the continuum Yang-Mills Hamiltonian to the discrete version).

The second term is the mass term, and the alternating sign is a signature of the staggered formulation. The last term accounts for the fermionic chemical potential. The third term in the Hamiltonian corresponds to the invariant Casimir operator of the theory and represents the color electric field energy stored in the gauge links. More precisely,  $\hat{\boldsymbol{L}}_{n}^{2}=\hat{L}_{n}^{a}\hat{L}_{n}^{a}=\hat{R}_{n}^{a}\hat{R}_{n}^{a}$ where $\hat{L}_{n}^{a}$  and $\hat{R}_{n}^{a}$ (with $a=x,y,z$) are respectively the left and right color electric field components on link $n$. They are conjugate momenta of the vector potential \cite{kogut_introduction_1979}. The operators $\hat{L}_{n}^{a}$ and $\hat{R}_{n}^{a}$ satisfy the algebra $[\hat{R}_{n}^{a},\hat{R}_{m}^{b}]=i\epsilon_{abc}\hat{R}_{n}^{c}\delta_{mn}$,  $[\hat{L}_{n}^{a},\hat{L}_{m}^{b}]=-i\epsilon_{abc}\hat{L}_{n}^{c}\delta_{mn}$, and $[\hat{L}_{n}^{a},\hat{R}_{m}^{b}]=0$, where $\epsilon_{abc}$ is the Levi-Civita symbol.
For a non-Abelian gauge group, the right and left color electric field are related via the adjoint representation $\hat{R}_{n}^{a}=(\hat{U}_n^\text{adj})_{ab}\hat{L}_{n}^{b}$, with  $(\hat{U}_{n}^\text{adj})_{ab}=2\mathrm{Tr}\left[ \hat{U}_{n}T^{a}\hat{U}_{n}^{\dagger}T^{b}\right].$

Due to gauge invariance, the Hamiltonian in equation (\ref{SIeq:KSham}) commutes with the Gauss' law operators (i.e the generators of local gauge transformation) 
\begin{equation}
\hat{G}_{n}^{a}\equiv \hat{L}_{n}^{a}-\hat{R}_{n-1}^{a}-\hat{Q}_{n}^{a}, \quad a=x,y,z,
\end{equation}
where $\hat{L}_{n}^{a}$ and $\hat{R}_{n-1}^{a}$ act on the links emanating from the site $n$, which itself carries the non-Abelian color charge $\hat{Q}_{n}^{a}=\phi_{n}^{i \dagger}(T^{a})_{ij}\phi_{n}^{j}$. In absence of static or external charges, the physical states $\ket{\Psi_{phys}}$ of the theory must obey the Gauss law  $\hat{G}_{n}^{a}\ket{\Psi_{phys}}=0$. In one spatial dimension and for open boundary conditions, the Gauss law can be used to integrate and eliminate the gauge fields to obtain a purely fermionic Hamiltonian. 
This approach was recently used in tensor network study of string breaking phenomena in non-Abelian lattice gauge theories \cite{sala_variational_2018}. The main idea behind this gauge elimination process is to seek a unitary transformation $\hat{\Theta}$ such that the connection term $\hat{U}_{n}$ in the kinetic energy disappears i.e. $\hat{\Theta} \left(\hat{\phi}_{n}^{\dagger}\hat{U}_{n}\hat{\phi}_{n+1}\right) \hat{\Theta}^{\dagger} =\hat{\phi}_{n}^{{\dagger}}\hat{\phi}_{n+1}$. The explicit expression for the transformation $\hat{\Theta}$ can be found in \cite{atas2021}. Here, we will only mention that it allows to express the Hamiltonian (\ref{SIeq:KSham}) exclusively in terms of fermionic degrees of freedom 

\begin{equation}
 \hat{H}_{\text{fermion}}\equiv\hat{\Theta} \hat{H} \hat{\Theta}^{\dagger}=\frac{1}{2} \sum_{n=1}^{N-1}\left( \hat{\phi}_{n}^{{\dagger}}\hat{\phi}_{n+1}+\mathrm{H.C.}\right) 
 +am\sum_{n=1}^{N}(-1)^{n}\hat{\phi}_{n}^{\dagger}\hat{\phi}_{n}+\frac{a^{2} g^{2}}{2} \hat{H}_{e} -a \mu \sum_{n=1}^{N} \hat{\phi}_{n}^\dagger \hat{\phi}_{n} , \label{SIeqrotatedHam}
\end{equation}
where we have multiplied the overall Hamiltonian by the lattice spacing $a$ to obtain a dimensionless one. In the following, we will adopt the conventional lattice units where $a = 1$. The color electric Hamiltonian takes the following form after the unitary transform 
\begin{equation}
    \hat{H}_{e}=\sum_{n=1}^{N-1}\left(\sum_{m\leq n} \hat{\mathbf{Q}}_{m}\right)^{2},
\end{equation}
where $\hat{\mathbf{Q}}_{m}$ is the vector of non-Abelian charges at site $m$ with components $Q_{m}^{a}$. Note that although the gauge fields do not appear explicitly, the non-Abelian physics is preserved in this formulation and reflected through the long
range exotic interaction between non-Abelian charges. Additionally, the physical states are now those which are compatible with the chosen boundary conditions, namely the color singlet must be the one with zero total non-Abelian charge $\hat{Q}_{tot}^{a}\ket{\Psi_{0}}=0$  with $\hat{Q}_{tot}^{a}=\sum_{n=1}^{N}\hat{Q}_{n}^{a}$ \cite{atas2021}. The global charge constraints appear in the expression of the projector in equation (\ref{eq:K_def_cartan}).

In order to study the model on a quantum computer, we perform a transformation on the fermionic Hamiltonian that allows us to write it in terms of qubits degree of freedom only. The transformation is achieved in two steps: first, the size of the lattice is doubled (tripled for SU(3)) and the colored fermionic fields are distributed among the new lattice sites by defining the single component fields $\hat{\psi}_{2n-1}=\hat{\phi}_{n}^{1}$, $\hat{\psi}_{2n}=\hat{\phi}_{n}^{2}$, $n=1,2,\dots,N$. Note that each site now hosts one fermion with a definite color, instead of two fermions of different color (see Fig.~\ref{fig:protocol}b). By construction, the new field is of fermionic nature as it just corresponds to a relabelling of the existing fermionic fields. 
In the second step, the single component fermionic field $\hat{\psi}_{n}$ is mapped into $\frac{1}{2}$-spin operators by means of a Jordan-Wigner transformation \cite{JordanWigner}
\begin{equation}
\hat{\psi}_{n}=\prod_{l<n}\left(-\hat{\sigma}_{l}^{z}\right)\hat{\sigma}_{n}^{-}, \quad \hat{\psi}_{n}^{\dagger}=\prod_{l<n}\left(-\hat{\sigma}_{l}^{z}\right)\hat{\sigma}_{n}^{+},
\end{equation}
where $\hat{\sigma}_{n}^{\pm}=(\hat{\sigma}_{n}^{x}\pm i \hat{\sigma}_{n}^{y})/2$. The string factor $\prod_{l<n}\left(-\hat{\sigma}_{l}^{z}\right)$ permits to recover the correct fermionic anticommutation relations for the field $\hat{\psi}_{n}$.
% \section{General description of SU$(2)$ lattice gauge theory}
% In the following, we present the Hamiltonian for an SU(2) LGT after the integration of the gauge field. 
Following the derivation in \cite{atas2021}, the qubit Hamiltonian can be written as 
\begin{equation}
	\hat{H} = \hat{H}_{kin}+m\hat{H}_{mass} + \frac{1}{2x}\hat{H}_{elec}-\mu \hat{H}_{chem}, \label{appendix:hamiltonian_general}
\end{equation}
where $\hat{H}_{kin}$ is the kinetic energy, $\hat{H}_{mass}$ is the mass term, $\hat{H}_{elec}$ is the color electric energy and $\hat{H}_{chem}$ is the chemical potential energy. Here $m$, $\mu$ and $x = 1/(g)^{2}$ are the dimensionless mass, chemical potential and coupling constant, respectively.
The qubit form of the different terms are given by
\begin{align}
	\hat{H}_{kin} &= -\frac{1}{2}\sum_{n=1}^{N-1}\left( \hat{\sigma}_{2n-1}^{+}\hat{\sigma}_{2n}^{z}\hat{\sigma}_{2n+1}^{-} + \hat{\sigma}_{2n}^{+}\hat{\sigma}_{2n+1}^{z} \hat{\sigma}_{2n+2}^{-} + \text{H.c.}\right),  \label{eq:qubitkinetic}\\
    \hat{H}_{mass} &= \sum_{n = 1}^{N} \left(\frac{(-1)^{n}}{2}\left( \hat{\sigma}_{2n-1}^{z}  + \hat{\sigma}_{2n}^{z}\right)+1 \right),\label{eq:qubitmass}\\
\notag	\hat{H}_{elec}&=\frac{3}{8}\sum_{n=1}^{N-1}(N-n)(1-\hat{\sigma}_{2n-1}^{z}\hat{\sigma}_{2n}^{z})\\  \notag &+\frac{1}{8}\sum_{n=1}^{N-2}\sum_{m>n}^{N-1}(N-m)\left(\hat{\sigma}_{2n-1}^{z}-\hat{\sigma}_{2n}^{z}\right)\left(\hat{\sigma}_{2m-1}^{z}-\hat{\sigma}_{2m}^{z}\right) \\
&+\sum_{n=1}^{N-2}\sum_{m>n}^{N-1}(N-m)\left(\hat{\sigma}_{2n-1}^{+}\hat{\sigma}_{2n}^{-}\hat{\sigma}_{2m}^{+}\hat{\sigma}_{2m-1}^{-}+\mathrm{H.c.}\right),\label{eq:qubitelectric} \\
\hat{H}_{chem}&=\frac{1}{4}\sum_{n=1}^{2N}\hat{\sigma}_{n}^{z},\label{eq:qubitchem}
\end{align}
where $\hat{\sigma}_{n}^{x,y,z}$ are the usual Pauli matrices and $N$ is the number of lattice sites.   As mentioned in the main text, the JW transformation induces four-body nonlocal interactions that appear in the expression of $\hat{H}_{elec}$. However, for a single unit cell (considered for the experiment), the nonlocal terms are absent.

$2N$ qubits are necessary to simulate our system since we have two colors in SU(2). The chemical potential term or baryon number is proportional to the total magnetisation of the system in the qubit formulation.
The model possesses a set of conserved charges given by \cite{atas2021}
\begin{align}
\hat{Q}_{tot}^{x}&=\frac{1}{2}\sum_{n=1}^{N}\left( \hat{\sigma}_{2n-1}^{+}\hat{\sigma}_{2n}^{-}+\mathrm{H.c.}\right), \label{appendix:su2_charges_x}\\
\hat{Q}_{tot}^{y}&=\frac{i}{2}\sum_{n=1}^{N}\left( \hat{\sigma}_{2n-1}^{-}\hat{\sigma}_{2n}^{+}-\mathrm{H.c.}\right), \label{appendix:su2_charges_y}\\
\hat{Q}_{tot}^{z}&=\frac{1}{4}\sum_{n=1}^{N}\left( \hat{\sigma}_{2n-1}^{z}-\hat{\sigma}_{2n}^{z}\right). \label{appendix:su2_charges_z}
 \end{align}
Physical states of the system must be color singlet i.e., they must be simultaneous zero modes of the total non-Abelian charges 
\begin{equation}
    \hat{Q}_{tot}^{a}\ket{\Psi}\equiv\sum_{n}\hat{Q}_{n}^{a}\ket{\Psi}=0, \quad a=x,y,z.
    \label{appendix:eq_zero_charge}
\end{equation}

\section{Qubit Hamiltonian for SU$(3)$ lattice gauge theory}
\label{sec:su3_general}
The derivation of the qubit Hamiltonian for SU(3) follows the same steps as for SU(2) with the exception that the fermion field appearing in equation (\ref{SIeq:KSham}) and (\ref{SIeqrotatedHam}) now has three color components $\hat{\phi}_{n}=(\hat{\phi}_{n}^{1},\hat{\phi}_{n}^{2},\phi_{n}^{3})^{T}$. The generators of the group also change and are now given by the Gell-Mann matrices $\hat{T}^{a}=\lambda^{a}/2$ with $a=1,2,\dots,8$. As a consequence, the non-Abelian charges must also be changed accordingly. The Hamiltonian is composed of four terms as in Eq.~(\ref{appendix:hamiltonian_general}). In the qubit formulation, the kinetic term is given by (see \cite{atas2023} for detailed derivation)
\begin{equation}
    \hat{H}_{kin}= \frac{1}{2}\sum_{n=1}^{N-1}(-1)^n\left(\hat{\sigma}_{3n-2}^{+}\hat{\sigma}_{3n-1}^{z}\hat{\sigma}_{3n}^{z} \hat{\sigma}_{3n+1}^{-}\right. 
    -\left. \hat{\sigma}_{3n-1}^{+}\hat{\sigma}_{3n}^{z}\hat{\sigma}_{3n+1}^{z} \hat{\sigma}_{3n+2}^{-}\right. 
    +\left. \hat{\sigma}_{3n}^{+}\hat{\sigma}_{3n+1}^{z}\hat{\sigma}_{3n+2}^{z} \hat{\sigma}_{3n+3}^{-} +\operatorname{H.c.}\right), \label{kinetic_ham_qubit}
\end{equation}
and the mass term reads
\begin{equation}
    \hat{H}_{mass}=\frac{1}{2}\sum_{n=1}^{N}\left[(-1)^{n}\left(\hat{\sigma}_{3n-2}^{z}+\hat{\sigma}_{3n-1}^{z}+\hat{\sigma}_{3n}^{z}\right)+3\right]. \label{mass_ham_qubit}
\end{equation}
The SU$(3)$ LGT possesses eight non-Abelian charges defined at each lattice sites as

\begin{align}
    \hat{Q}_{n}^{1}&=\frac{(-1)^{n}}{2}\left( \hat{\sigma}_{3n-2}^{+}\hat{\sigma}_{3n-1}^{-}+\operatorname{H.c.}\right)\label{qubit_nonabeliancharge1}, \\
    \hat{Q}_{n}^{2}&=\frac{i(-1)^{n}}{2}\left( \hat{\sigma}_{3n-1}^{+}\hat{\sigma}_{3n-2}^{-}-\operatorname{H.c.}\right)\label{qubit_nonabeliancharge2},\\
    \hat{Q}_{n}^{3}&=\frac{1}{4}\left( \hat{\sigma}_{3n-2}^{z}-\hat{\sigma}_{3n-1}^{z}\right)\label{qubit_nonabeliancharge3},\\
    \hat{Q}_{n}^{4}&=-\frac{1}{2}\left( \hat{\sigma}_{3n-2}^{+}\hat{\sigma}_{3n-1}^{z}\hat{\sigma}_{3n}^{-}+\operatorname{H.c.}\right)\label{qubit_nonabeliancharge4}, \\
    \hat{Q}_{n}^{5}&=\frac{i}{2}\left( \hat{\sigma}_{3n-2}^{+}\hat{\sigma}_{3n-1}^{z}\hat{\sigma}_{3n}^{-}-\operatorname{H.c.}\right)\label{qubit_nonabeliancharge5}, \\
    \hat{Q}_{n}^{6}&=\frac{(-1)^n}{2}\left( \hat{\sigma}_{3n-1}^{+}\hat{\sigma}_{3n}^{-}+\operatorname{H.c.}\right)\label{qubit_nonabeliancharge6},\\
    \hat{Q}_{n}^{7}&=\frac{i(-1)^n}{2}\left( \hat{\sigma}_{3n}^{+}\hat{\sigma}_{3n-1}^{-}-\operatorname{H.c.}\right)\label{qubit_nonabeliancharge7},\\
    \hat{Q}_{n}^{8}&=\frac{1}{4\sqrt{3}}\left( \hat{\sigma}_{3n-2}^{z}+\hat{\sigma}_{3n-1}^{z}-2\hat{\sigma}_{3n}^{z}\right)\label{qubit_nonabeliancharge8}.
\end{align}
% \begin{align}
%     \hat{Q}_{n}^{1}&=\frac{(-1)^{n}}{2}\left( \hat{\sigma}_{3n-2}^{+}\hat{\sigma}_{3n-1}^{-}+\operatorname{H.c.}\right)\label{qubit_nonabeliancharge1}, \\
%     \hat{Q}_{n}^{2}&=\frac{i(-1)^{n}}{2}\left( \hat{\sigma}_{3n-1}^{+}\hat{\sigma}_{3n-2}^{-}-\operatorname{H.c.}\right)\label{qubit_nonabeliancharge2},\\
%     \hat{Q}_{n}^{3}&=\frac{1}{4}\left( \hat{\sigma}_{3n-2}^{z}-\hat{\sigma}_{3n-1}^{z}\right)\label{qubit_nonabeliancharge3},\\
%     \hat{Q}_{n}^{4}&=-\frac{1}{2}\left( \hat{\sigma}_{3n-2}^{+}\hat{\sigma}_{3n-1}^{z}\hat{\sigma}_{3n}^{-}+\operatorname{H.c.}\right)\label{qubit_nonabeliancharge4}, \\
%     \hat{Q}_{n}^{5}&=\frac{i}{2}\left( \hat{\sigma}_{3n-2}^{+}\hat{\sigma}_{3n-1}^{z}\hat{\sigma}_{3n}^{-}-\operatorname{H.c.}\right)\label{qubit_nonabeliancharge5}, \\
%     \hat{Q}_{n}^{6}&=\frac{(-1)^n}{2}\left( \hat{\sigma}_{3n-1}^{+}\hat{\sigma}_{3n}^{-}+\operatorname{H.c.}\right)\label{qubit_nonabeliancharge6}, \\
%     \hat{Q}_{n}^{7}&=\frac{i(-1)^n}{2}\left( \hat{\sigma}_{3n}^{+}\hat{\sigma}_{3n-1}^{-}-\operatorname{H.c.}\right)\label{qubit_nonabeliancharge7},\\
%     \hat{Q}_{n}^{8}&=\frac{1}{4\sqrt{3}}\left( \hat{\sigma}_{3n-2}^{z}+\hat{\sigma}_{3n-1}^{z}-2\hat{\sigma}_{3n}^{z}\right)\label{qubit_nonabeliancharge8}.
% \end{align}
Note that there are two diagonal charges $\hat{Q}_{n}^{3}$ and $\hat{Q}_{n}^{8}$ at each site.
The color electric field Hamiltonian can be obtained by using the qubit expressions for the non-Abelian charges in Eqs.~(\ref{qubit_nonabeliancharge1}-\ref{qubit_nonabeliancharge8}) and reads
\begin{align}
    \hat{H}_{elec}=&\notag \frac{1}{3}\sum_{n=1}^{N-1}(N-n) \\
   & \notag\times  
     \left( 3-\hat{\sigma}_{3n-2}^{z}\hat{\sigma}_{3n-1}^{z}-\hat{\sigma}_{3n-2}^{z}\hat{\sigma}_{3n}^{z}-\hat{\sigma}_{3n-1}^{z}\hat{\sigma}_{3n}^{z}\right)\\\notag &+\sum_{n=1}^{N-2}\sum_{m=n+1}^{N-1}\left[ (N-m)\left( \hat{\sigma}_{3n-2}^{+}\hat{\sigma}_{3n-1}^{-}\hat{\sigma}_{3m-1}^{+}\hat{\sigma}_{3m-2}^{-}\right. \right. \\
   \notag & +\left. \hat{\sigma}_{3n-1}^{+}\hat{\sigma}_{3n}^{-}\hat{\sigma}_{3m-1}^{-}\hat{\sigma}_{3m}^{+}+\operatorname{H.c.}\right)(-1)^{n+m} \\
   \notag & +(N-m)\left(\hat{\sigma}_{3n-2}^{+}\hat{\sigma}_{3n-1}^{z}\hat{\sigma}_{3n}^{-}\hat{\sigma}_{3m-2}^{-}\hat{\sigma}_{3m-1}^{z}\hat{\sigma}_{3m}^{+} +\operatorname{H.c.}\right) \\
   \notag &-\frac{1}{12}(N-m) \hat{\sigma}_{3m-2}^{z}(\hat{\sigma}_{3n-1}^{z}+\hat{\sigma}_{3n}^{z}-2\hat{\sigma}_{3n-2}^{z}) \\
   \notag &-\frac{1}{12}(N-m) \hat{\sigma}_{3m-1}^{z}(\hat{\sigma}_{3n}^{z}+\hat{\sigma}_{3n-2}^{z}-2\hat{\sigma}_{3n-1}^{z})\\
    & \left. -\frac{1}{12}(N-m) \hat{\sigma}_{3m}^{z}(\hat{\sigma}_{3n-2}^{z}+\hat{\sigma}_{3n-1}^{z}-2\hat{\sigma}_{3n}^{z}) \right], \label{qubit_electric_general}
\end{align}
which exhibits long range qubit-qubit interaction as a direct consequence of the gauge elimination \cite{atas2023}. However, for a single unit cell, these terms do not appear in the Hamiltonian. The chemical potential operator is proportional to the total magnetization of the system in the qubit encoding 
\begin{equation}
    \hat{H}_{chem}=\frac{1}{6}\sum_{n=1}^{3N}\hat{\sigma}_{n}^{z}. \label{qubit_baryon_number}
\end{equation}
Note that one needs $3N$ qubits to simulate the system for SU(3) due to the three available color degrees of freedom. 

\section{Circuit decomposition and transpilation in terms of native gates}\label{app:circ_decomp} 

\begin{figure}[!ht]
    \centering
    \includegraphics[width=0.5\linewidth]{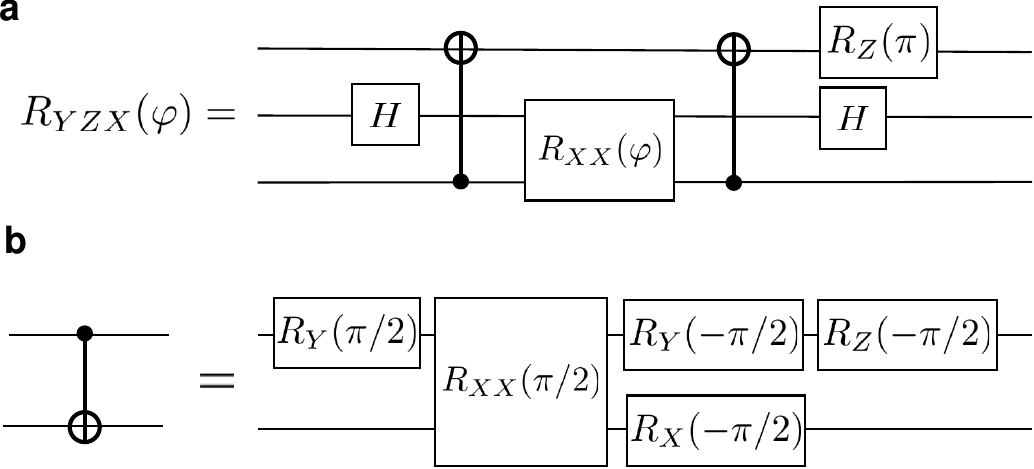}
    \caption{(a) Decomposition of the three-body gate $R_{YZX}(\varphi)$ in terms of CNOT gates, Hadamard $H$ gates and entangling $R_{XX}$ gates. (b) Decomposition of the CNOT gate in terms of the native trapped ion entangling M\o lmer-S\o rensen gate and local rotations. }
    \label{fig:three_body_gate_decomp}
\end{figure}

\begin{figure}[ht]
    \centering
    \includegraphics[width=0.5\linewidth]{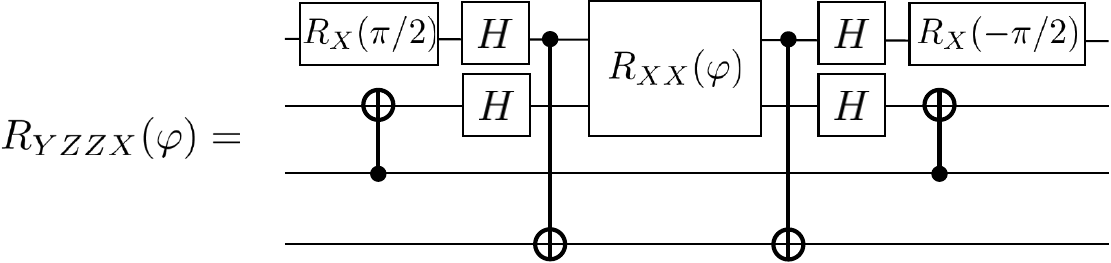}
    \caption{Decomposition of the four-body gate $R_{YZZX}(\varphi)$ in terms of CNOT gates and entangling $R_{XX}$ gate. The circuit can be further simplified and expressed in native gates by using the decomposition of the CNOT gate in Fig.~\ref{fig:three_body_gate_decomp}(b).}
    \label{fig:four_body_gate_decomp}
\end{figure}

In this section, we discuss the circuit decomposition and transpilation into native gates of the multi-qubit gates used in our SU$(2)$ and SU$(3)$ quantum circuits (see Fig.~\ref{fig:SU2_circuit} and \ref{fig:methods_su3_circuit}). In Fig.~\ref{fig:three_body_gate_decomp}b, we show the decomposition of the CNOT gates into MS gates, the native two-qubit operations on our ion trap platform. Fig.~\ref{fig:three_body_gate_decomp}a shows the decomposition of the three-body gate used in the SU$(2)$ circuit. In total, three MS gates are necessary to implement the three-body $R_{YZX}$ gate.  Similarly, Fig.~\ref{fig:four_body_gate_decomp} illustrates the decomposition of the four-body gate $R_{YZZX}$ into native gates.

\begin{figure*}[ht]
    \centering
    \includegraphics[width=0.83\linewidth]{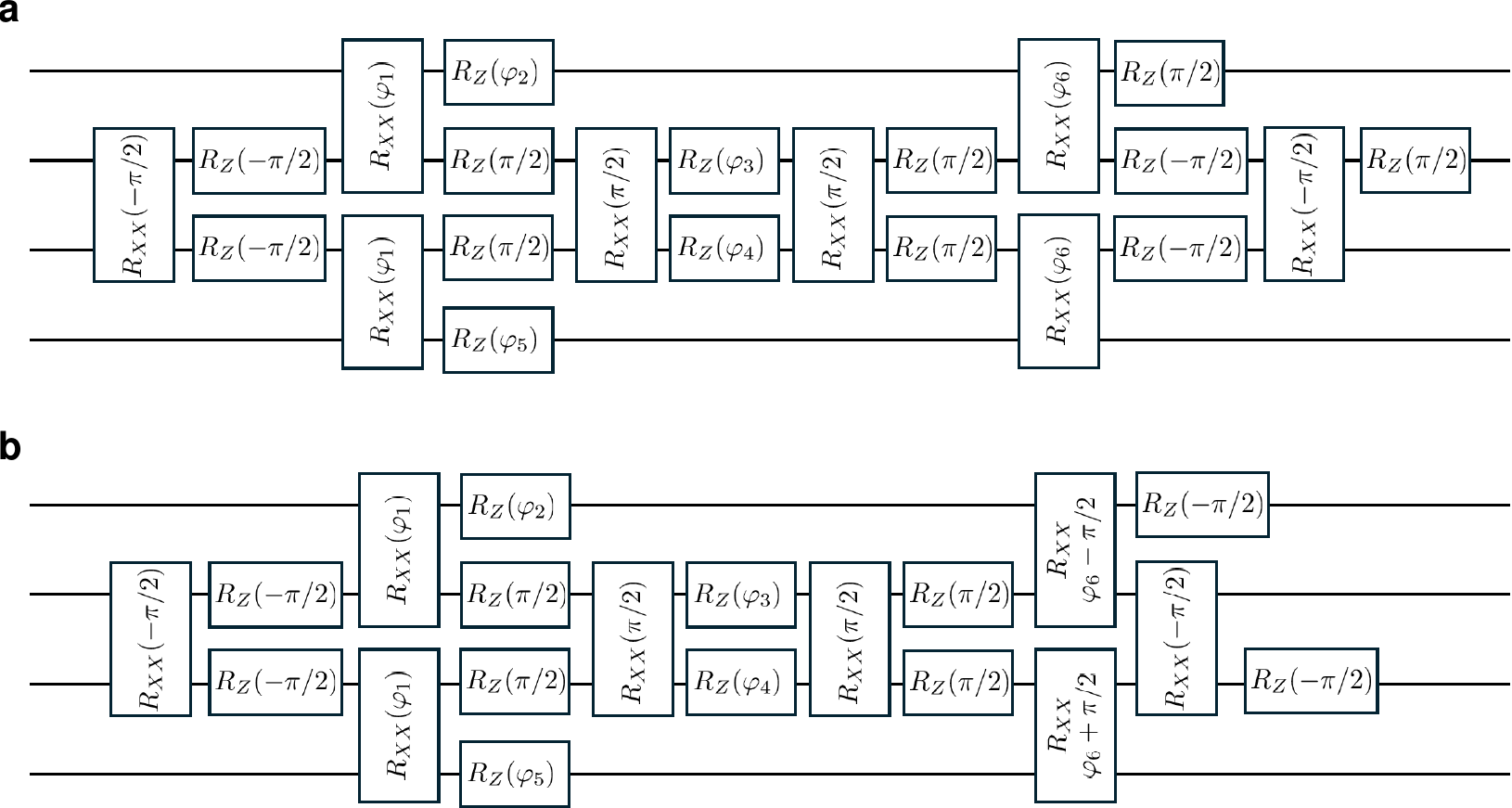}
    \caption{(a) Reduced system circuit used for measuring $\hat{H}_1$, i.e. the diagonal part of the Hamiltonian  of SU$(2)$ LGT unit cell on the trapped-ion device. It contains 6 variational parameters and 8 entangling MS gates. (b) The measurement circuit $\hat{M}_H$ to transform the non-diagonal Pauli strings into diagonal Pauli strings is combined with the system ansatz $\hat{U}_S(\boldsymbol{\varphi})$. This combination leaves the total number of entangling gates the same, i.e., adding the measurement circuit does not increase the number of MS gates in the circuit. Both (a) and (b) are also used for noisy VQE simulations.}
    \label{fig:reduced_su2_circ}
\end{figure*}

The two circuits used in our SU$(2)$ experiment and VQE simulations are shown in Fig.~\ref{fig:reduced_su2_circ}. 
The structure of the circuit in Fig.~\ref{fig:SU2_circuit}a allows for  simplifications when the four $R_{YZX}$ are replaced with their decomposition in MS gates, leading to the circuit in Fig~\ref{fig:reduced_su2_circ}a. This circuit contains eight MS gates in total and is used to measure the diagonal part of the Hamiltonian $\hat{H}_1$ of SU(2) LGT unit
cell on the trapped-ion device.  To measure the non diagonal part of the Hamiltonian $\hat{H}_2$, we append the measurement circuit $\hat{M}_{H}$ shown at the end of the system circuit in Fig.~\ref{fig:SU2_circuit}. Although the measurement circuit contains six CNOT gates, the number of entangling gates stays the same as for the diagonal part due to gate cancellations and simplifications (see Fig.~\ref{fig:reduced_su2_circ}b).  

\begin{figure*}[!ht]
    \centering
    \includegraphics[width=0.98\linewidth]{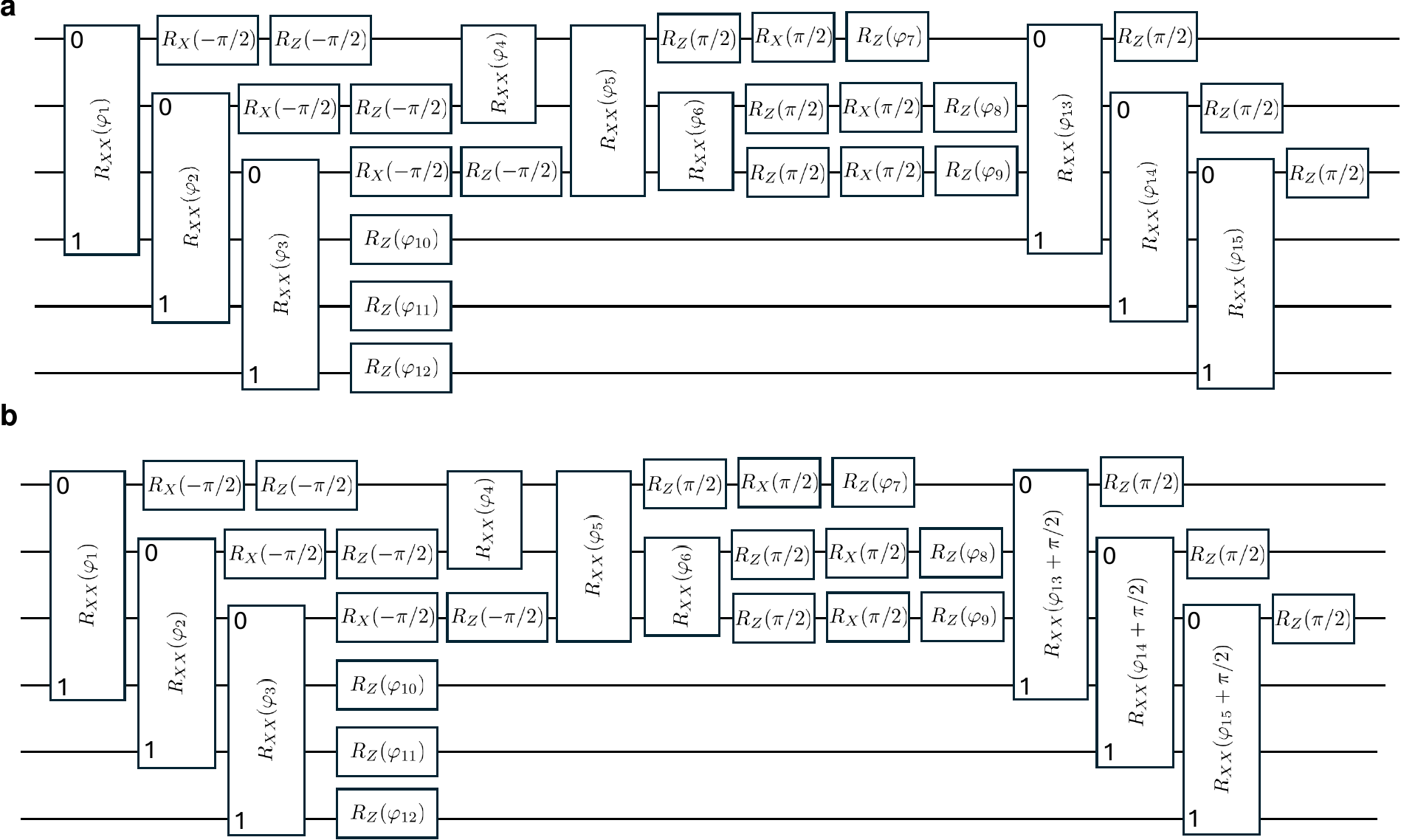}
    \caption{(a) Reduced system circuit for SU$(3)$ that is used in the experiment to measure the diagonal family of Pauli strings and the order parameter. The $0$ and $1$ in the $R_{XX}$ gates indicate which qubits are entangled. The circuit contains 15 variational parameters. (b) When the measurement circuit $\hat{M}_H$, designed to measure the non-diagonal energy contribution $\hat{H}_2$, is combined with the circuit in (a), the total number of MS gates remains unchanged. Only the parameters of the last three MS gates are shifted by $\pi/2$ as a result of merging $\hat{M}_H$ with (a).  These circuits are also used for our noisy VQE simulations. }
    \label{fig:reduced_su3_circ}
\end{figure*}

For SU(3), the four-body gate $R_{YZZX}$ is decomposed into entangling gates using the circuit in Fig.~\ref{fig:four_body_gate_decomp}. By using gate identities, we can reduce the MS gate count for both circuit needed for the measurement of $\hat{H}_{1}$ (diagonal) and $\hat{H}_{2}$ (non-diagonal). The reduced circuit used in our SU(3) experiment and our numerical simulations are shown in Fig.~\ref{fig:reduced_su3_circ} and involve only 9 MS gates. 

\clearpage
\putbib[biblio]
\end{bibunit}

%\clearpage
%\bibliographystylesupp{plain}
%\bibliographysupp{biblioSI}

\end{document}